\documentclass[twoside,11pt]{article}

\usepackage[accepted]{melba}

\usepackage{amsmath,amsfonts}

\usepackage{graphicx}
\usepackage{tabularray}
\usepackage{hyperref}
\usepackage{float}
\usepackage{listings}
\usepackage{courier}

\usepackage[capitalize]{cleveref}
\crefname{section}{Sec.}{Secs.}
\Crefname{section}{Section}{Sections}
\Crefname{table}{Table}{Tables}
\crefname{table}{Tab.}{Tabs.}

\Crefname{equation}{Equation}{Equations}
\crefname{equation}{Eq.}{Eqs.}

\usepackage{authblk}

\melbaid{2023:002}  
\melbaauthors{Kofler et al.}
\volume{2}
\firstpageno{27}  
\year{2023}  
\datesubmitted{05/2022}  
\datepublished{05/2023}  

\ShortHeadings{Are we using appropriate segmentation metrics?}{Kofler et al.}

\title{Are we using appropriate segmentation metrics? \\ Identifying correlates of human expert perception for CNN training beyond rolling the DICE coefficient}

\author[1,2,3,12]{Florian Kofler}
\author[1,2]{Ivan Ezhov}
\author[4,5]{Fabian Isensee}
\author[6]{Fabian Balsiger}
\author[1]{Christoph Berger}
\author[1]{Maximilian Koerner}
\author[7]{Beatrice Demiray}
\author[7,8]{Julia Rackerseder}
\author[1,9,10]{Johannes Paetzold}
\author[1,11]{Hongwei Li}
\author[1,2]{Suprosanna Shit}
\author[6]{Richard McKinley}
\author[12]{Marie Piraud }
\author[13,14,15]{Spyridon Bakas}
\author[3]{Claus Zimmer}
\author[7]{Nassir Navab}
\author[3]{Jan Kirschke}
\author[2,3,16]{Benedikt Wiestler}
\author[1,11,16]{Bjoern Menze}

\affil[1]{Department of Informatics, Technical University Munich, Germany }
\affil[2]{TranslaTUM - Central Institute for Translational Cancer Research, Technical University of Munich, Germany}
\affil[3]{Department of Diagnostic and Interventional Neuroradiology, School of Medicine, Klinikum rechts der Isar, Technical University of Munich, Germany}
\affil[4]{Applied Computer Vision Lab, Helmholtz Imaging, Germany}
\affil[5]{Division of Medical Image Computing, German Cancer Research Center (DKFZ), Germany}
\affil[6]{Support Center for Advanced Neuroimaging (SCAN), Institute for Diagnostic and Interventional Neuroradiology, Inselspital, Bern University Hospital, University of Bern, Bern, Switzerland}
\affil[7]{Computer Aided Medical Procedures (CAMP), Technical University of Munich, Germany}
\affil[8]{ImFusion GmbH, Munich, Germany}
\affil[9]{Helmholtz Zentrum München, Germany}
\affil[10]{Imperial College London, Exhibition Rd, South Kensington, London SW7 2BX, United Kingdom}
\affil[11]{Department of Quantitative Biomedicine, University of Zurich, Switzerland}
\affil[12]{Helmholtz AI, Helmholtz Zentrum München, Germany}
\affil[13]{Center for Biomedical Image Computing and Analytics (CBICA), University of Pennsylvania, Philadelphia, Pennsylvania, USA }
\affil[14]{Department of Pathology and Laboratory Medicine, Perelman School of Medicine, University of Pennsylvania, Philadelphia, Pennsylvania, USA}
\affil[15]{Department of Radiology, Perelman School of Medicine, University of Pennsylvania, Philadelphia, Pennsylvania, USA}
\affil[16]{contributed equally as senior authors}

\begin{document}

\maketitle

\newpage
\begin{abstract}
Metrics optimized in complex machine learning tasks are often selected in an ad-hoc manner.
It is unknown how they align with human expert perception.
We explore the correlations between established quantitative segmentation quality metrics and qualitative evaluations by professionally trained human raters.
Therefore, we conduct psychophysical experiments for two complex biomedical semantic segmentation problems.
We discover that current standard metrics and loss functions correlate only moderately with the segmentation quality assessment of experts.
Importantly, this effect is particularly pronounced for clinically relevant structures, such as the enhancing tumor compartment of glioma in brain magnetic resonance and grey matter in ultrasound imaging.
It is often unclear how to optimize abstract metrics, such as human expert perception, in convolutional neural network (CNN) training.
To cope with this challenge, we propose a novel strategy employing techniques of classical statistics to create complementary compound loss functions to better approximate human expert perception.
Across all rating experiments, human experts consistently scored computer-generated segmentations better than the human-curated reference labels.
Our results, therefore, strongly question many current practices in medical image segmentation and provide meaningful cues for future research.
\end{abstract}



\section{Introduction}
In the medical imaging community, challenges have become a prominent forum to benchmark the latest methodological advances for complex machine learning problems \citep{maier2018rankings}.
Across many semantic segmentation challenges \citep{menze2014multimodal,sekuboyina2020verse,bilic2019liver,bakas2018identifying}, convolutional neural networks (CNNs), and in particular the U-Net architecture \citep{ronneberger2015u}, gained increasing popularity over the years.
Most challenges rely upon a seemingly ad-hoc combination of Dice coefficient (DICE) with other metrics for scoring \citep{menze2014multimodal, maier2018rankings}.

How these metrics and the resulting rankings reflect clinical relevance and expert opinion regarding segmentation quality is poorly understood.
For example, segmentation models in the BraTS challenge \citep{bakas2018identifying} are evaluated  on three label channels: \emph{enhancing tumor (ET)}, \emph{tumor core (TC)} and \emph{whole tumor (WT)} compartments.
Volumetric Dice coefficients and Hausdorff distances are aggregated equally across the three challenges to obtain the final challenge ranking.
In contrast, the ET channel is of higher importance from a medical perspective.
An increase in ET volume defines tumor progression \citep{wen2010updated}, and glioma surgery aims for a gross total resection of the ET \citep{weller2014eano}.

\citet{taha2015metrics} investigated how different metrics capture certain aspects of segmentation quality.
However, there is a clear gap in knowledge of approximating expert assessment using established metrics and achieving a clinically meaningful representation of segmentation performance.
This problem is of central importance when defining loss functions for CNN training; consequently, the optimized metrics are often selected in an ad-hoc fashion.
In contrast to a plethora of volumetric loss functions \citep{hashemi2018asymmetric,milletari2016v,sudre2017generalised,rahman2016optimizing,brosch2015deep,salehi2017tversky}, only a few non-volumetric losses are established for CNN training such as \citet{karimi2019reducing}\footnote{This is especially true when searching for multi-class segmentation losses, which support GPU-based training. For instance, according to the authors, there is no implementation of \citet{karimi2019reducing} with GPU support.}.

We aim to identify the quantitative correlates of human expert perception concerning semantic segmentation quality assessment.
Therefore, we conduct psychophysical segmentation rating experiments on multi-modal 3D magnetic resonance (MR) glioma data and 2D ultrasound (US) data for grey- and white matter segmentation with expert radiologists.

Building upon these insights, we propose a method exploiting techniques of classical statistics and experimental psychology to form new compound losses for convolutional neural network(CNN) training.
Even though our CNN training experiments do not achieve the initially intended goal, we derive meaningful insights for future research in semantic segmentation.
Across our quality assessment experiments, human experts consistently score computer-generated segmentations better than human-curated reference labels.
This finding raises many questions about the status quo of semantic segmentation in medical images.

 \section{Methodology}
\subsection{Collection of expert ratings}
\label{text:rating_collection}
\begin{figure*}[htb]
\centering
\includegraphics[width=0.95\textwidth]{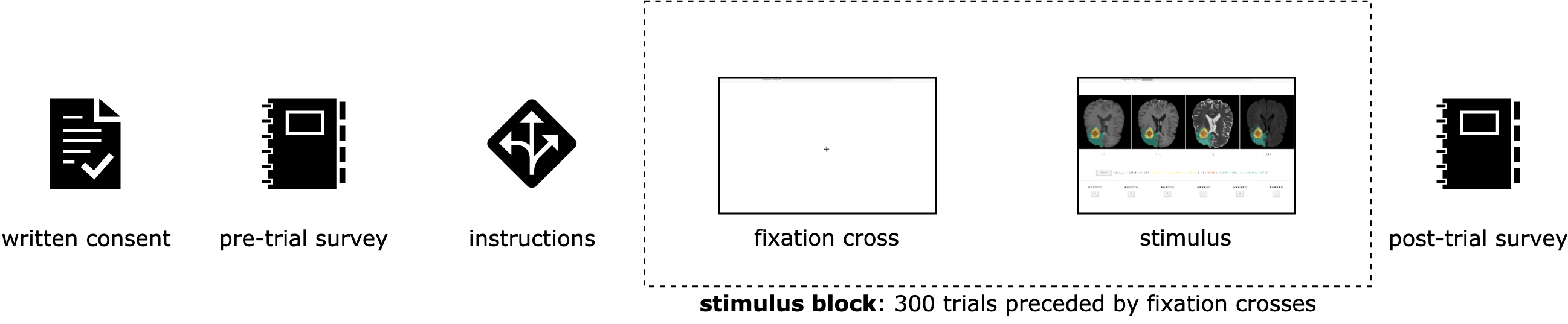}
\caption{\small{Chronological sequence of the experiments from left to right.
Participants conduct the evaluations in a suitable environment for reading medical exams.
The experiments start with participants declaring consent.
Afterward, participants answer a survey regarding their age, gender, and various items to measure their expertise.
Subsequently, the stimulus trials are presented in random sequence to account for order effects.
Following the rating assignment, the experiment automatically progresses to the presentation of the next trial.
At each experiment's end, experts have the opportunity to provide feedback during a post-trial survey.
}}
\label{fig:experiment}
\end{figure*}

To better understand how experts judge the quality of medical segmentations, we conduct experiments where participants rate segmentation quality on a six-degree Likert scale (see \Cref{fig:experiment}).
The selection options range from one star for \emph{strong rejection} to six stars for \emph{perfect} segmentations.
Participants assign star ratings by pressing the corresponding keys on the keyboard.
The hotkeys are selected, accounting for several international keyboard layouts to minimize and equalize reaction times.
All stimuli appear on a neutral-grey background.
The segmentation labels and other colorful items are presented employing a color-blind-friendly color palette \citep{wong2011points}.
Participants could toggle the display of the semi-opaque segmentation labels by pressing the space bar.
A fixation cross appears in a quasi-random inter-trial interval (ITI) of \emph{125 / 250 / 500 / 750} milliseconds to ensure participants cannot anticipate the onset of subsequent trials.
We recorded how often participants toggled the display of the segmentation labels and reaction times for all tasks.
By comparing the individual ratings against the mean, the methodology enables analyzing the participants' individual biases (see Appendix).
To allow the recording of accurate reaction times, the experiment was realized using \emph{JS Psych v6.1} \citep{de2015jspsych} embedded into a \emph{Vue JS v2.6}  web application.

\subsection{Computation of metrics}
We calculate a comprehensive set of segmentation quality metrics for each experimental condition using \emph{pymia} \citep{jungo2021pymia}, summarized in \Cref{tab:metrics_implementations}.
The metrics compare the segmentations to the expert-curated reference images.

\begin{table}[hbt]
\caption{Metrics considered in the analysis.
The metrics are categorized according to \citet{taha2015metrics}.}
    \centering
    \scriptsize
    \begin{tblr}{colspec={lcr},colspec={Q[1.5,c] Q[1,c] Q[2,c] Q[1.5,c]}}
    Metric & Abbreviation & Reference &
    Category \\
    \hline

    Dice coefficient &
    DICE &
    \citep{dice1945measures} &
    overlap \\

    Jaccard coefficient &
    JACRD &
    \citep{jaccard1912distribution} &
    overlap \\

    Sensitivity &
    SNSVTY &
    \citep{tharwat2020classification} &
    overlap \\

    Specificity &
    SPCFTY &
    \citep{tharwat2020classification} &
    overlap \\

    Fallout &
    FALLOUT &
    \citep{tharwat2020classification} &
    overlap \\

    False negative rate &
    FNR &
    \citep{tharwat2020classification} &
    overlap \\

    Accuracy &
    ACURCY &
    \citep{tharwat2020classification} &
    overlap \\

    Precision &
    PRCISON &
    \citep{tharwat2020classification} &
    overlap \\

    False positive &
    FP &
    \citep{pearson1904theory} &
    overlap \\
    
    True positive &
    TP &
    \citep{pearson1904theory} &
    overlap \\
    
    False negative &
    FN &
    \citep{pearson1904theory} &
    overlap \\
    
    True negative &
    TN &
    \citep{pearson1904theory} &
    overlap \\

    F-measure &
    FMEASR &
    \citep{tharwat2020classification} &
    overlap \\

    Volume similarity &
    VOLSMTY &
    \citep{cardenes2009multidimensional} &
    volume \\

    Prediction volume &
    PREDVOL &
    \citep{kohlberger2012evaluating} &
    volume \\
    
    Reference volume &
    REVOL &
    \citep{kohlberger2012evaluating} &
    volume \\

    Global consistency error  &
    GCOERR &
    \citep{martin2001database} &
    misc \\

    Rand index &
    RNDIND &
    \citep{rand1971objective} &
    pair-counting \\

    Adjusted rand index &
    ADJRIND &
    \citep{hubert1985comparing} &
    pair-counting \\

    Mutual information &
    MUTINF &
    \citep{cover1999elements} &
    information-theoretical \\

    Variation of information &
    VARINFO &
    \citep{meilua2003comparing} &
    information-theoretical \\

    Interclass correlation &
    ICCORR &
    \citep{shrout1979intraclass} &
    probabilistic \\
   
    Probabilistic distance &
    PROBDST &
    \citep{gerig2001valmet} &
    probabilistic \\

    Cohen Kappa coefficient &
    KAPPA &
    \citep{cohen1960coefficient} &
    probabilistic \\

    Area under curve &
    AUC &
    \citep{powers2020evaluation} &
    probabilistic \\

    Hausdorff distance &
    HDRFDST &
    \citep{huttenlocher1993comparing} &
    distance \\

    Average distance &
    AVGDIST &
    \citep{fujita2013metrics} &
    distance \\

    Mahalanobis distance &
    MAHLNBS &
    \citep{mahalanobis1936generalized} &
    distance \\

    Surface Dice overlap &
    SURFDICE &
    \citep{nikolov2018deep} &
    distance \\
    
    Surface overlap &
    SURFOVLP &
    \citep{nikolov2018deep} &
    distance \\

    \end{tblr}
    \label{tab:metrics_implementations}
\end{table}

\subsection{Loss function implementations}
\Cref{tab:loss_implementations} summarizes the loss functions implementations employed in our experiments.
For many losses, multiple implementations exist.
We discovered that multiple implementations of the same losses could produce different loss values.
Implementations might arise from different means of approximation, such as the choice of $\epsilon$.
To investigate whether the choice of implementation plays a role, we included multiple implementations in the analysis.
For Tversky loss, we choose combinations of hyperparameters that have proven successful for similar segmentation problems; in the following analysis, the parameters for $\alpha$ and $\beta$ are written behind the abbreviation.

\begin{table}[hbt]
    \caption{Losses considered in our analysis. Implementations are Github links.}
    \centering
    \scriptsize
    \begin{tblr}{colspec={lcr},colspec={Q[1.5,c] Q[1,c] Q[1,c] Q[2,c]}}
    Loss name & Abbreviation & Implementation & Reference \\
    \hline
    Asymmetric         & ASYM         & \href{https://github.com/JunMa11/SegLoss/blob/71b14900e91ea9405d9705c95b451fc819f24c70/losses_pytorch/dice_loss.py\#L390}{junma}           & \cite{hashemi2018asymmetric}           \\
    Binary cross-entropy    & BCE   & \href{https://github.com/pytorch/pytorch/blob/159c48b19bce9e7a2d95e308a8c7e87327062563/torch/nn/modules/loss.py\#L616}{pytorch}              & n/a           \\
    (Soft) Dice         & DICE         & \href{https://github.com/Project-MONAI/MONAI/blob/4b8819aa2d26f96dc2d23d3159e56cc413d3e846/monai/losses/tversky.py\#L22}{monai}            & \cite{milletari2016v}           \\
    Soft Dice         & SOFTD         & \href{https://github.com/MIC-DKFZ/nnUNet/blob/05c9f323752252913bf975e09cfe73e92b38d335/nnunet/training/loss_functions/dice_loss.py\#L158}{nnUNet}            & \cite{drozdzal2016importance}           \\
    Generalized Dice         & GDICE\_L         & \href{https://github.com/LIVIAETS/boundary-loss/blob/108bd9892adca476e6cdf424124bc6268707498e/losses.py\#L29}{liviaets}              & \cite{sudre2017generalised}           \\
    Generalized Dice         & GDICE\_W         & \href{https://github.com/wolny/pytorch-3dunet/blob/6e5a24b6438f8c631289c10638a17dea14d42051/unet3d/losses.py\#L75}{wolny}             & \cite{sudre2017generalised}           \\
    Generalized Dice         & GDICE\_M         & \href{https://github.com/Project-MONAI/MONAI/blob/4b8819aa2d26f96dc2d23d3159e56cc413d3e846/monai/losses/dice.py\#L25}{monai}            & \cite{sudre2017generalised}           \\
    Hausdorff DT         & HDDT         & \href{https://github.com/PatRyg99/HausdorffLoss/blob/9f580acd421af648e74b45d46555ccb7a876c27c/hausdorff_loss.py\#L80}{patryg}             & \cite{karimi2019reducing}           \\
    Hausdorff ER         & HDER         & \href{https://github.com/PatRyg99/HausdorffLoss/blob/9f580acd421af648e74b45d46555ccb7a876c27c/hausdorff_loss.py\#L80}{patryg}             & \cite{karimi2019reducing}           \\
    Jaccard         & IOU         & \href{https://github.com/JunMa11/SegLoss/blob/71b14900e91ea9405d9705c95b451fc819f24c70/losses_pytorch/dice_loss.py\#L293}{junma}             & \cite{rahman2016optimizing}           \\
    Jaccard         & JAC         & \href{https://github.com/Project-MONAI/MONAI/blob/4b8819aa2d26f96dc2d23d3159e56cc413d3e846/monai/losses/dice.py\#L25}{monai}             & \cite{milletari2016v}           \\
    Sensitivity-Specifity         & SS         & \href{https://github.com/JunMa11/SegLoss/blob/71b14900e91ea9405d9705c95b451fc819f24c70/losses_pytorch/dice_loss.py\#L192}{junma}             & \cite{brosch2015deep}           \\
    Tversky         & TVERSKY         & \href{https://github.com/Project-MONAI/MONAI/blob/4b8819aa2d26f96dc2d23d3159e56cc413d3e846/monai/losses/tversky.py\#L22}{monai}            & \cite{salehi2017tversky}           \\
    \end{tblr}
    \label{tab:loss_implementations}
\end{table}

\subsection{Statistical analysis}
\label{text:statistical_analysis}
All statistical analyses are computed with \emph{R}.
We implement linear mixed models via \emph{lme4} \citep{bateslme} and evaluate their performance using the package \emph{performance} \citep{performance}.
The principal component analysis is illustrated with \emph{factorminer} \citep{factominer}.

\subsection{Construction of compound losses}
\label{text:loss_construction}
To achieve a better correlation with human segmentation quality assessment, we aim to design new loss functions by combining established loss functions in a complementary fashion.
The resulting compound loss functions are constructed as a weighted linear combination of established loss functions per label channel:

\begin{equation}
L  = \sum_{i} \alpha_i \sum_{j} w_{ij} loss_{ij}
\label{eq:compound}
\end{equation}

where $\alpha_i$ denotes weights for channel $i$, and $w_{ij}$ denotes weights for loss $j$ in channel $i$.
Here, the model coefficients obtained from the linear mixed models serve as weights.

Further information regarding the mixed model experiments can be found in \Cref{text:losses_derived} and \Cref{text:appendix_lmms}.

\section{Experiments}
\subsection*{Experiment 1: MR segmentation rating}
\label{text:brats_experiment}
\noindent \textbf{Motivation:}
To understand how human experts evaluate segmentation quality, we conduct a perception study.
We select the BraTS 2019 test set  \citep{menze2014multimodal} as a platform for our experiments for several reasons.
First, glioma segmentation represents a complex multi-class segmentation problem.
Second, the annotations have been curated by multiple domain experts over the years.
Consequently, unlike in other available segmentation data sets, the annotations not only represent the interpretation of a single individual and are of high quality.
Last, the dataset is non-public, so we can be sure that the evaluated algorithms are not trained on the annotations.

\begin{figure*}[hbtp]
    \centering
    \includegraphics[width=0.8\textwidth]{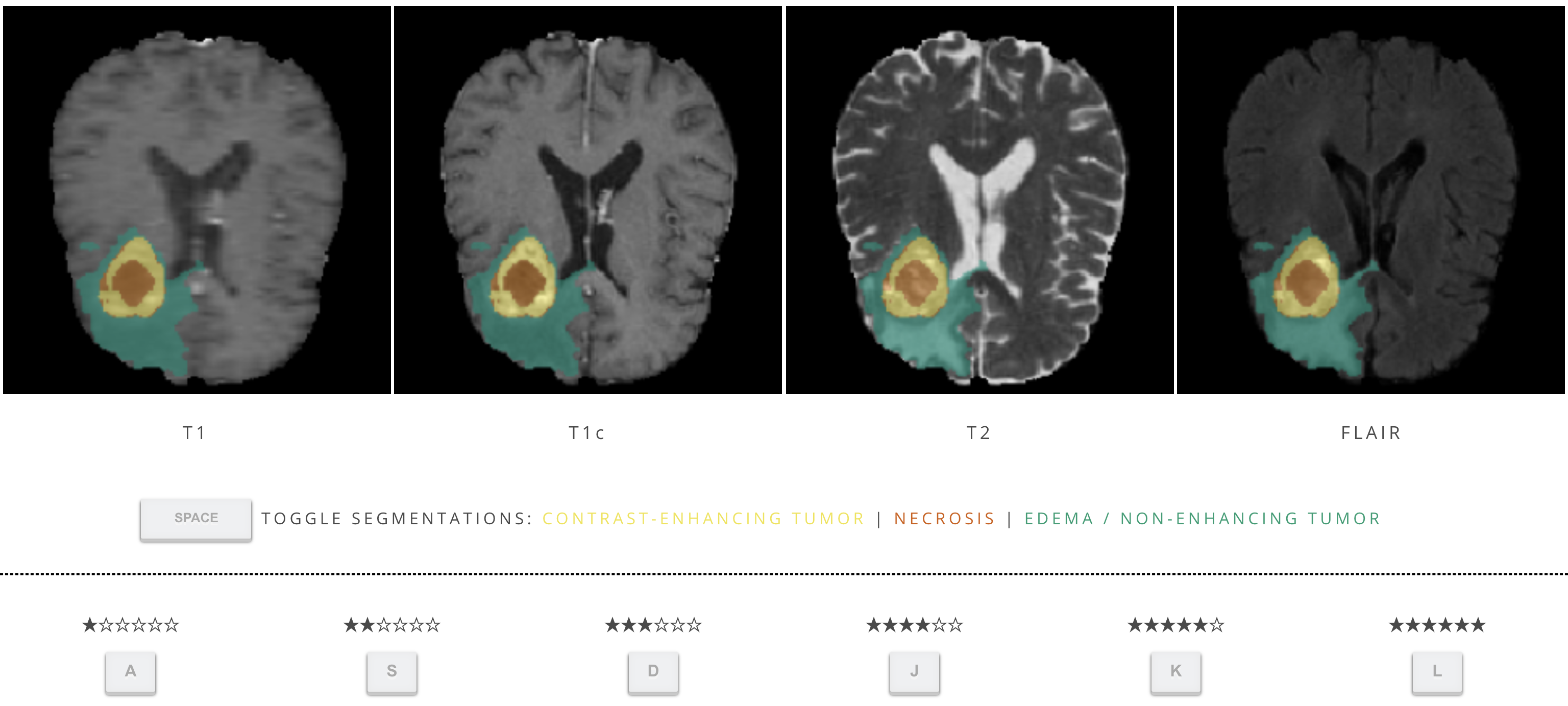}
    \caption{\small{Stimulus material and rating controls presented to the participants in the BraTS segmentation rating experiment}.
    One trial consisted of the presentation of an MR exam in either axial, sagittal, or coronal view, along with the controls for the quality rating and a legend for the segmentation labels.
    We presented the glioma's center of mass according to the TC, defined by the \emph{necrosis and enhancing tumor} voxels in 2D slices of T1, T1c, T2, T2 FLAIR in a horizontal row arrangement.
    The stimulus presentation is conducted in line with best practices in experimental psychology; further details are outlined in \Cref{fig:experiment} and in \Cref{text:rating_collection}}
    \label{fig:stimuli}
\end{figure*}

\noindent \textbf{Procedure:} 
We randomly select three exams from eight institutions that provided data to the BraTS 2019 test set \citep{menze2014multimodal}.
Additionally, we present one exam with apparently erroneous segmentation labels to check whether participants follow the instructions.
We display stimulus material according to four experimental conditions, i.e. four different segmentation labels for each exam: the human-annotated \emph{reference} segmentations, segmentations from team \emph{zyx} \citep{zhao2019multi}, a \emph{SIMPLE} \citep{langerak2010label} fusion of seven segmentations \citep{zhao2019multi,xfeng,isensee2018no,scan,lfb,gbmnet,econib} obtained from BraTS-Toolkit \citep{kofler2020brats}, and a segmentation randomly selected without replacement from all the teams participating in BraTS 2019.
In order to have an unbiased view, we select the non-public BraTS test set for this experiment, as we can be sure that none of the evaluated algorithms is trained on it.
For all \textbf{25} exams, the respective center of mass is displayed on \textbf{3} axes, namely axial, coronal, and sagittal view for each of the \textbf{4} experimental conditions.
This results in a total of \textbf{25 * 3 * 4 = 300} trials presented to each expert radiologist.
 \Cref{fig:stimuli} depicts an example trial.

Furthermore, we calculate metrics comparing to the official reference labels for all BraTS evaluation criteria: \emph{WT}, \emph{TC}, and \emph{ET}, as well as the individual labels: \emph{necrosis} and \emph{edema}.
In addition, we compute mean aggregates for BraTS and individual labels.

\begin{figure}[hbtp]
    \centering
    \includegraphics[width=0.95\textwidth]{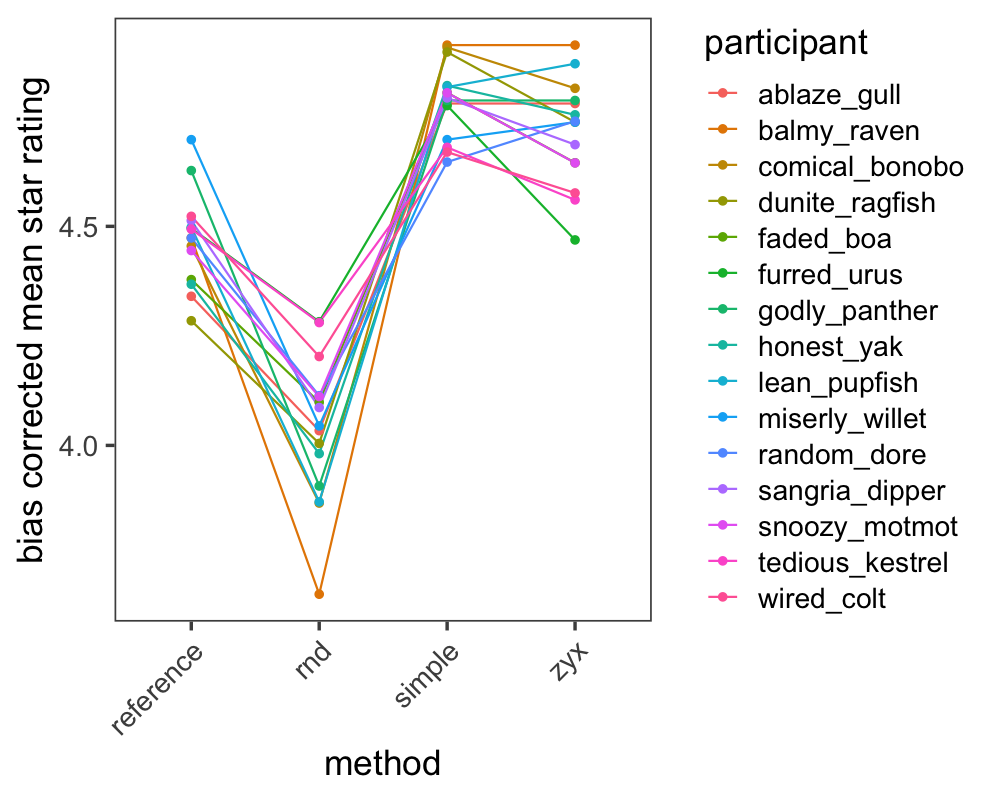}
    \caption{\small{
    Bias-corrected mean star ratings per expert and method for the MR segmentation rating experiment.
    Raw ratings and the bias correction methodology are detailed in Appendix.
    Notably, the human-annotated reference labels (\emph{gt}) are predominantly rated worse than computer-generated annotations of \emph{simple} and \emph{zyx}.
    }} \label{fig:brats_bc}
\end{figure}

\begin{figure}[H]
    \centering
    \includegraphics[width=0.95\textwidth]{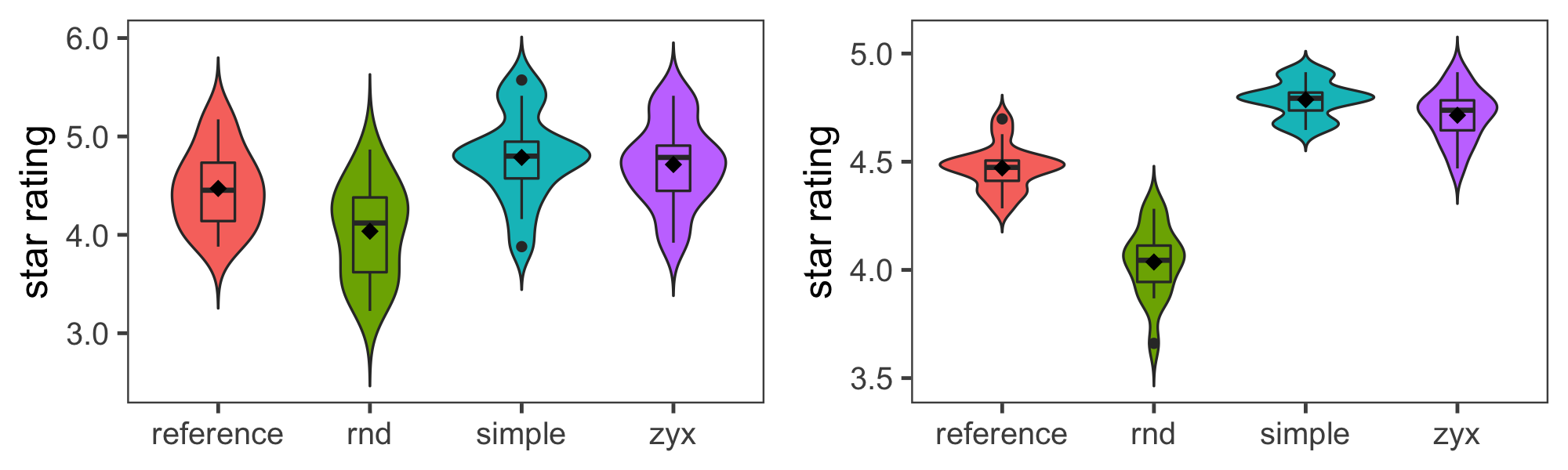}
    \caption{\small{Mean expert assessment in star rating across exams for the different experimental conditions.
    Ratings \emph{without} (left) vs. ratings \emph{with} bias correction (right).
    Diamonds indicate mean scores.
    Expert radiologists rated the \emph{simple} fusion with a mean of \emph{4.79} stars.
    This is slightly higher than the best individual docker algorithm \emph{zyx}} with \emph{4.71} stars.
    The fusions' mean improvement is mainly driven by more robust segmentation performance with fewer outliers towards the bottom.
    We observe that both of these algorithmic segmentations receive slightly higher scores than the human expert-annotated \emph{reference} labels at \emph{4.47} stars.
    The quality of the randomly selected BraTS submissions \emph{(rnd)} still lags behind with \emph{4.04} stars.
    Three more qualitative segmentation rating examples can be found in the Appendix.
    } \label{fig:brats_stars}
\end{figure}

\begin{figure*}[hbt]
    \centering
    \includegraphics[width=1.0\linewidth]{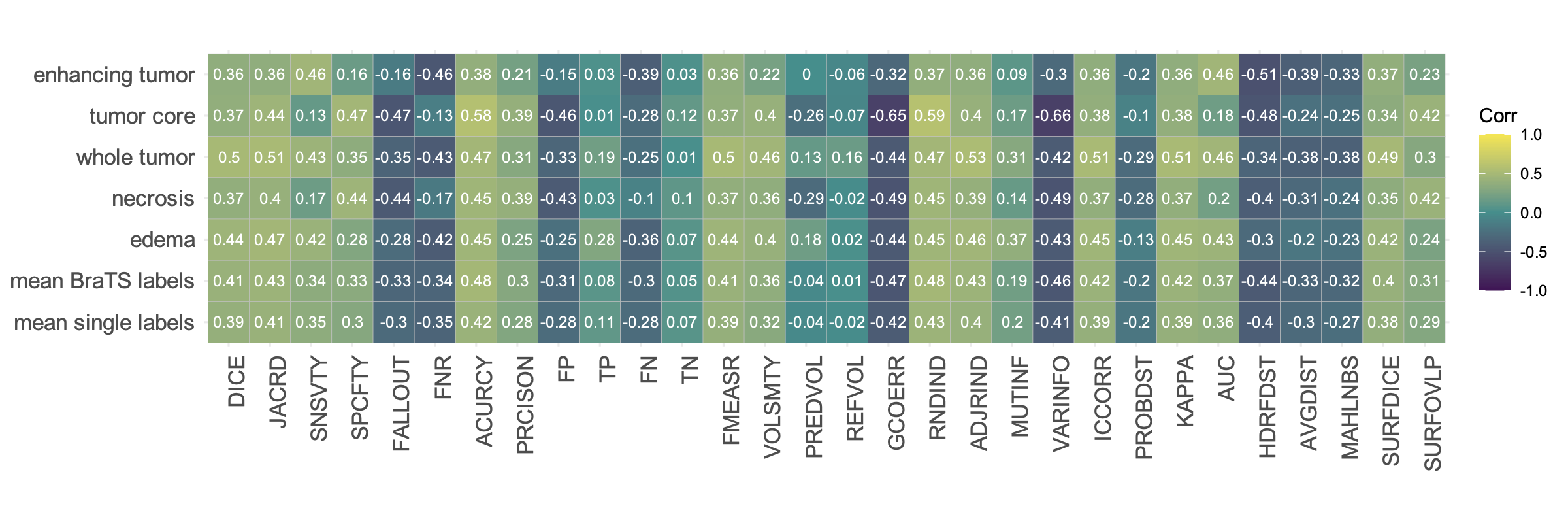}
    \caption{\small{Pearson correlation matrix between expert assessment and segmentation quality metrics.  The rows show the correlations for the individual label channels. In addition, we present the mean aggregates for the BraTS labels (\emph{enhancing tumor, tumor core, and whole tumor}), as well as the single channels (\emph{enhancing tumor, necrosis, and edema}). For the abbreviations in the figure refer to \citep{taha2015metrics,jungo2021pymia}.}}
    \label{fig:metrics}
\end{figure*}

\noindent \textbf{Results:}
A total of \emph{n=15} radiologists (\emph{two} female, \emph{13} male) from \emph{six} institutions participated in the experiment.
Participants had an average experience of 10.0$\pm$5.1 working years as radiologists and a mean age of 37.7$\pm$4.8.
Uninterrupted, participants needed approximately 50 minutes to complete the experiment.
 \Cref{fig:brats_bc} and \Cref{fig:brats_stars} depict how experts evaluated the segmentation quality.
 \Cref{fig:metrics} depicts the Pearson correlation matrix between segmentation quality metrics and expert assessment.
We find only low to moderate correlations across all metrics; correlations are especially low for the clinically important \emph{enhancing tumor} label.
It is important to note that the frequently used DICE coefficient is outperformed by other less established metrics.

\subsection*{Experiment 2: US segmentation rating}
\noindent \textbf{Motivation:}
To understand how the findings from Experiment 1 translate into other segmentation problems, we conduct another expert perception study.
We investigate the task of tissue segmentation in ultrasound (US) imaging in a dataset from \citet{demiray2019weaklysupervised}.
The dataset has the interesting property that the reference annotations are created in a multi-stage collaborative human-machine interaction:
\citet{demiray2019weaklysupervised} base their work on the RESECT dataset \citep{xiao2017re}.
In addition to co-registered pre-operative T1-weighted and T2-weighted MR images and the intra-operative US images from different stages of the intervention, RESECT provides corresponding anatomic landmarks for both modalities.
First, the co-registered pre-operative MR and pre-durotomy US are re-registered with an affine registration with LC2 metric \citep{fuerst2014automatic} in ImFusion Suite \footnote{ImFusion GmbH, Munich, Germany, https://www.imfusion.com}.
Second, two human experts selected the registration result with smaller registration errors according to the RESECT landmarks as a starting point for manual corrections of the registration.
Third, the tissues in the MR volumes are segmented with FreeSurfer \citep{fischl2012freesurfer} and manually corrected by two medical imaging experts (work experience $>$3 years).
Fourth, the tissue segmentation maps are transformed from the MR space into the US space.
Last, the human experts again perform quality control on the resulting annotations.

\begin{figure}[H]
    \centering
    \includegraphics[width=0.95\textwidth]{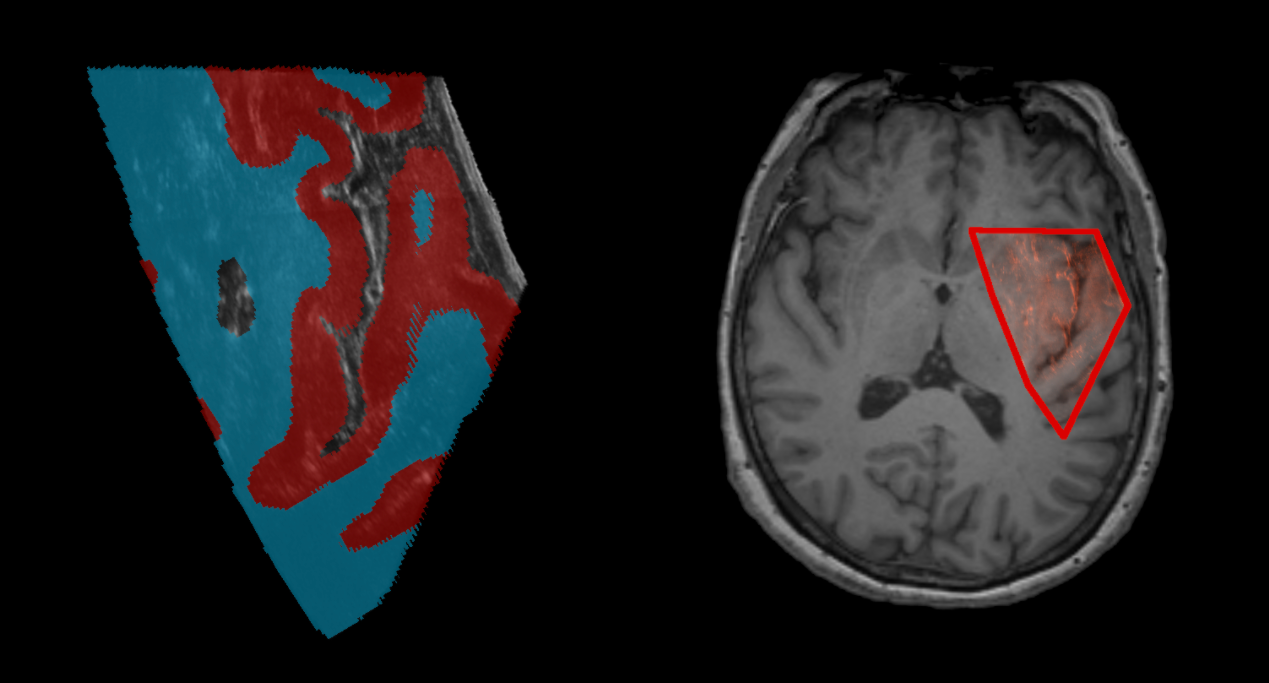}
    \caption{\small{Stimulus material presented to the participants during the US segmentation rating experiment.
    One trial consisted of the presentation of a US image (left) with the source area delineated in a corresponding MR image (right).
    In addition, the star rating assessment control elements were displayed analog to the BraTS experiment, see \Cref{fig:stimuli}.
    As in the previous experiment, participants could toggle the display of segmentation maps for grey- and white matter with the space bar.
    Further details regarding the stimulus presentation are outlined in Section \Cref{text:rating_collection}}}
    \label{fig:us_stimuli}
\end{figure}

\noindent \textbf{Procedure:}
For the perception study, human experts evaluate the performance of different tissue segmentation algorithms against the reference.
Analogue to the first experiment, the experts are blinded regarding the source of the segmentation.
In addition to the reference annotation, we present two segmentation algorithms from Demiray et al. (in preparation).
First, a segmentation algorithm based on DeepMedic \citep{kamnitsas2017efficient} and another based on nnUnet \citep{isensee2021nnu}.
Further, a control condition with broken annotations is presented to the participants.
 \Cref{fig:us_stimuli} illustrates one example of the total 57 trials.
In line with Experiment 1, we compute segmentation quality metrics between the two candidate algorithms, the control condition, and the reference labels and correlate them with expert assessment.

\begin{figure}[H]
    \centering
    \includegraphics[width=0.95\textwidth]{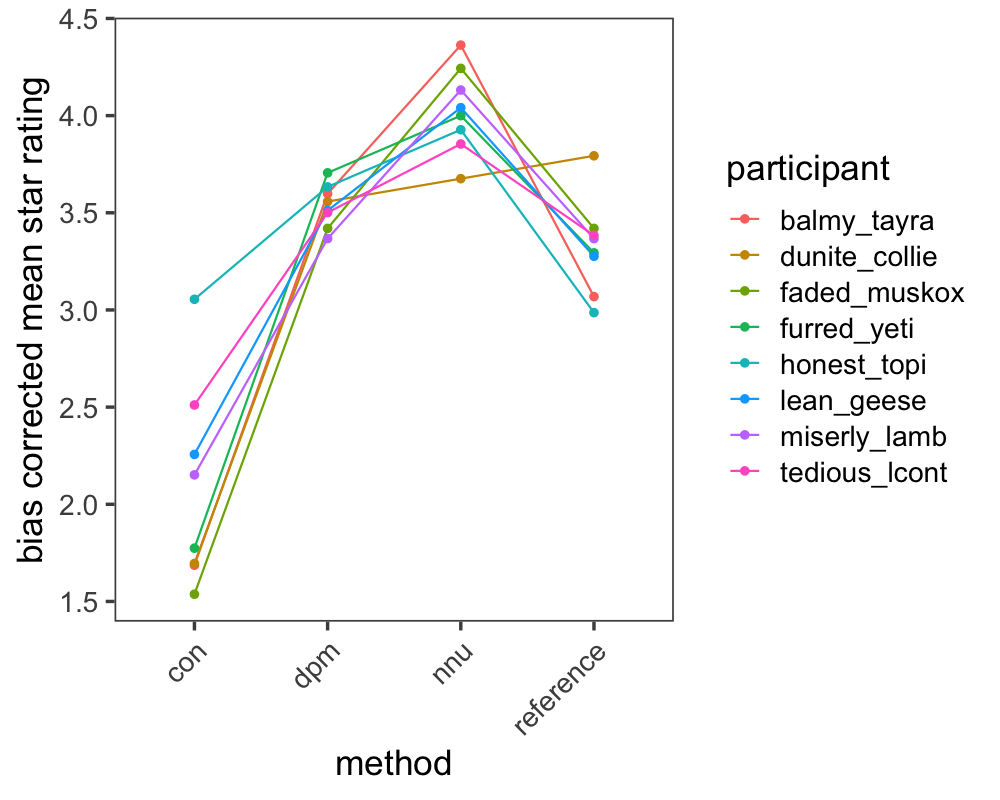}
    \caption{\small{
    Bias-corrected mean star ratings per expert and condition for the ultrasound segmentation rating experiment.
    Expert radiologists rated the \emph{nnu} (inspired by \citep{isensee2021nnu}) and \emph{dpm} (inspired by \citep{kamnitsas2017efficient,kamnitsas2015multi}) candidate algorithms much higher than the control condition (\emph{con}) consisting out of purposely wrongly manufactured segmentations.
    Notably, again the U-Net segmentation achieves the highest ratings.
    }}
    \label{fig:us_bc}
\end{figure}

\noindent \textbf{Results:}
A total of \emph{n=8} radiologists (\emph{three} female, \emph{five} male), from \emph{two} institutions participated in the experiment.
Participants had an average work experience of 5.25$\pm$5.63 years and a mean age of 35.0$\pm$5.42.
Uninterrupted, participants needed around twenty minutes to complete the experiment.
 \Cref{fig:us_bc} depicts how experts evaluated the segmentation quality across experimental conditions.

\begin{figure*}[htb]
    \centering
    \includegraphics[width=1.0\linewidth]{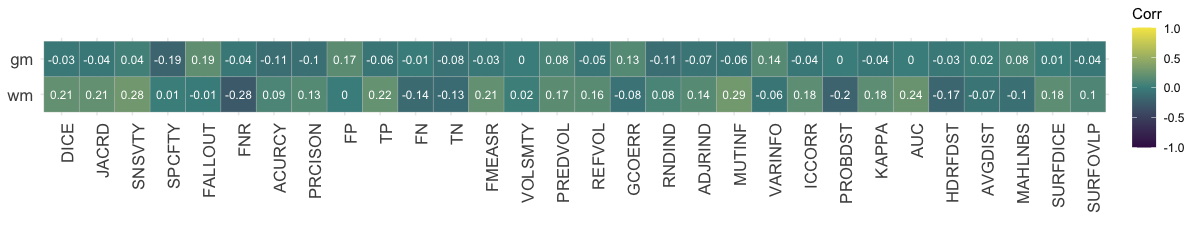}
    \caption{\small{Pearson correlation matrix between expert assessment and segmentation quality metrics for the US segmentation rating experiment.
    The rows show the correlations for grey matter (\emph{gm}) and white matter (\emph{wm}).
    }} \label{fig:us_metrics}
\end{figure*}

Furthermore, \Cref{fig:us_metrics} depicts the Pearson correlation matrix between segmentation quality metrics and the expert assessment.
Similar to the first Experiment, we find only low correlations across all metrics, especially for the grey matter label.

\subsection*{Experiment 3: Evaluation of established loss functions}
\noindent \textbf{Motivation:}
To evaluate how these findings could translate into CNN training.
A differentiable loss function is required to train a CNN with stochastic gradient descent (SGD).
As all metrics do not fulfill this criterion, we investigate how established loss functions correlate with expert quality assessment.

\noindent \textbf{Procedure:}
To achieve this, we use the binary label maps to compute established segmentation losses (defined in \Cref{tab:loss_implementations}) between the candidate segmentations and the \emph{reference} labels for the BraTS segmentation task.
To test the validity of this approach, we conduct a supportive experiment, comparing \emph{binary} segmentations to \emph{non-binary} network outputs and achieve comparable results across all loss functions.
In line with the previous experiments, we then correlate the resulting losses with expert assessment.
Furthermore, we conduct a principal component analysis (PCA) to better understand loss functions' complementarity.

\begin{figure*}[hbt]
    \centering
    \includegraphics[width=1.0\linewidth]{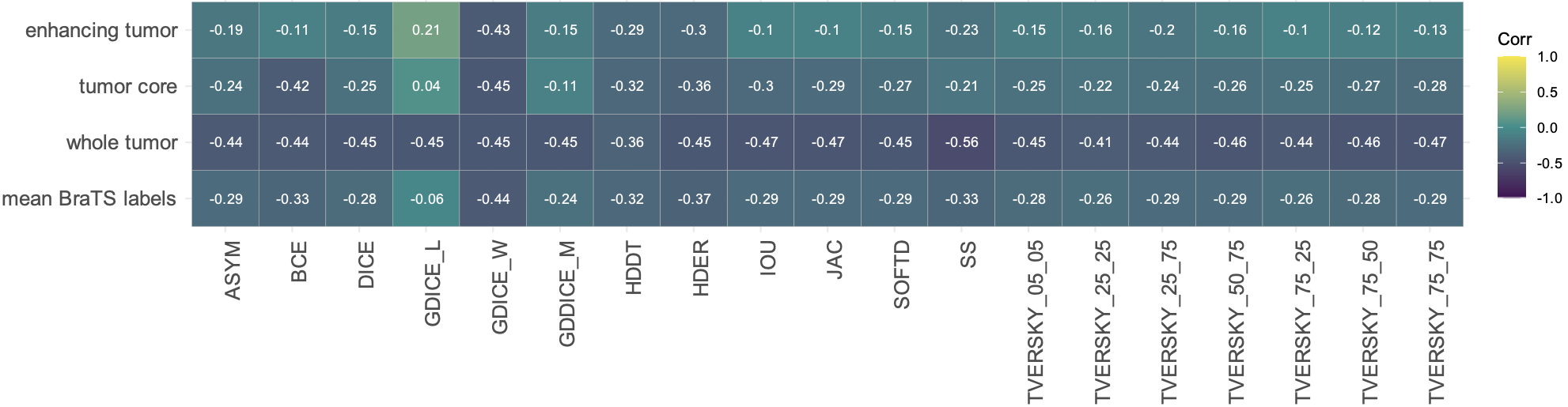}
    \caption{\small{Pearson correlation matrix between expert assessment and segmentation losses. With the exception of one Generalized Dice Score implementation, we observe only low correlations for the enhancing tumor and tumor core channel. In contrast, multiple losses are moderately correlated for the whole tumor channel. We observe huge variance across the implementations of \emph{Generalized Dice}. However, the implementations abbreviated as IOU and JAC for \emph{Jaccard index}, as well as DICE and SOFTD for \emph{Dice coefficient}, provide very similar signals, as expected.}}
    \label{fig:losses}
\end{figure*}

\noindent \textbf{Results:}
\Cref{fig:losses} illustrates the correlations with expert assessment; again, we only find low to moderate correlations.
Furthermore, the results of the PCA are depicted in the Appendix.

\subsection*{Experiment 4: Construction of compound loss functions}
\label{text:losses_derived}
\noindent \textbf{Motivation:}
As all investigated established loss functions achieve, at best moderate correlation with expert assessment, we set out to explore whether we can achieve a better fit with expert assessment when combining these.

\begin{figure}[H]
    \label{fig:loss_map}
    \centering
    \includegraphics[width=0.95\textwidth]{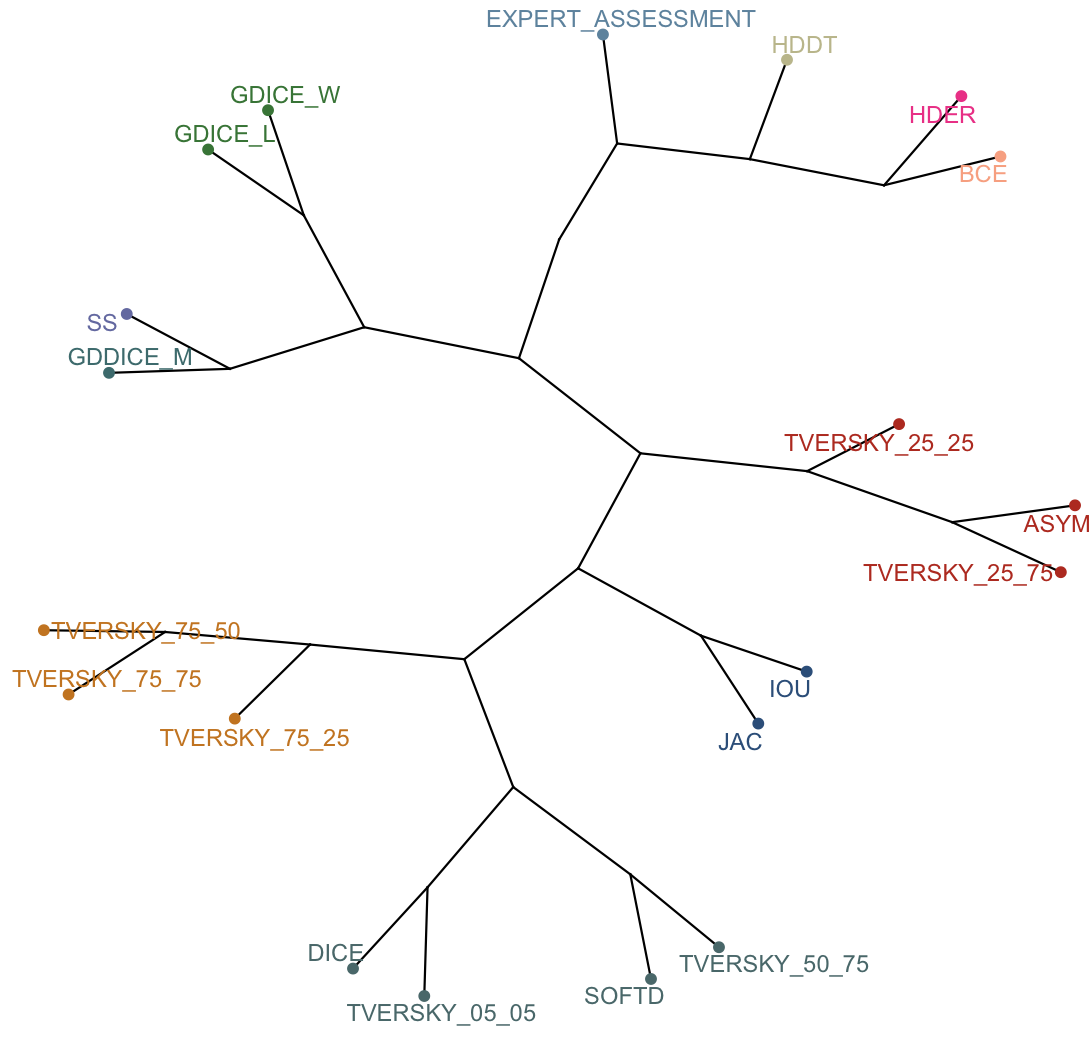}
    \caption{\small{Phylogenetic tree illustrating hierarchical clustering on a Euclidean distance matrix.
    Losses are colored according to ten cluster groups.
    We included the expert assessment for reference; one can observe how it resides somewhere between the distance-based losses and the group of generalized Dice losses and SS loss.
    These findings are in line with the \emph{PCA}, see Appendix.}}
\end{figure}

\noindent \textbf{Procedure:}
We approach this in a two-step process.
First, we identify promising loss combinations:
Losses produce different signals (gradients) when they react to input signals (network outputs).
With hierarchical clustering on a Euclidean distance matrix, we analyze similarity in loss response patterns, see \Cref{fig:loss_map}.

Applying \Cref{eq:compound}, we can now build complementary loss combinations by selecting established loss functions from the different and hopefully complementary cluster groups.

In a second step, we evaluate our loss combinations' predictive performance for the expert assessment using linear mixed models (LMM) as detailed in section \Cref{text:statistical_analysis}.
Therefore, we average the human expert rating across views to obtain a \emph{quasi-metric} variable allowing us to apply mixed models.
We deem this approach valid, as the distribution is consistent across views; see Appendix.
We then model the human expert assessment as a dependent variable and predict it by plugging in the loss values of our candidate combinations as fixed-effect predictor variables.
Mixed models allow accounting for the \emph{non-independence} of data points by additionally modeling random factors.
Therefore, we include \emph{exam} as a random factor, as some glioma are more difficult to segment.
This increased difficulty is expressed in less segmentation performance and greater variability across algorithmic and human segmentations.
Additionally we include the \emph{segmentation method} as a random factor, as some algorithms are better than others and human segmentations are worse than good algorithms according to our experts, compare \Cref{text:brats_experiment}.
To identify promising compound loss candidates for CNN training, we evaluate the predictive power of our models by computing Pseudo R\textsuperscript{2} \citep{nakagawa2017coefficient} while monitoring typical mixed regression model criteria, such as multi-collinearity, non-normality of residuals and outliers, homoscedasticity, homogeneity of variance and normality of random effects, see \Cref{text:appendix_lmms}.
Once a promising loss combination is detected, we obtain the weights for our loss formulation (see \Cref{text:loss_construction}) from the corresponding LMM.

\noindent \textbf{Results:}
Our methodology allows for generating a variety of loss candidates.
To evaluate the performance of our approach, we obtain four promising loss candidates.
The first two candidates use the same combination across all label channels.
Candidate \emph{gdice\_bce} uses \emph{GDICE\_W} in combination with \emph{BCE}.
Candidate \emph{gdice\_ss\_bce} iterates upon this by adding \emph{SS} loss to the equation.
In contrast the candidates \emph{channel\_wise\_weighted} and \emph{channel\_wise} use different losses per channel:
Channel \emph{whole tumor} uses the \emph{gdice\_ss\_bce} candidate loss, while channel \emph{tumor core} is computed via the \emph{gdice\_bce} variant.
In contrast, the \emph{enhancing tumor} channel relies solely on \emph{GDICE\_W}, as this is the only loss candidate with at least moderate correlation with expert assessment, compare \Cref{fig:losses}.
While candidate \emph{channel\_wise} treats all channels equally, the weighted variant prioritizes the more clinically relevant \emph{tumor core and enhancing tumor channels} by a factor of \emph{five}.
Detailed analysis for all models is provided in the Appendix.

\subsection*{Experiment 5: Quantitative evaluation of compound losses}
\noindent \textbf{Motivation:}
To explore how the four identified compound losses from Experiment 4 perform with regard to established metrics, we perform another experiment.

\noindent \textbf{Procedure:}
Therefore, we train nnU-Net \citep{isensee2021nnu} using a standard BraTS trainer \citep{isensee2020nnu} with a moderate augmentation pipeline on the \emph{fold 0 split} of the \emph{BraTS 2020} training set, using the rest of the training set for validation. The official \emph{BraTS 2020} validation set is used for testing.
To make our experiment fully reproducible, we select the BraTS 2020 data for this experiment as the training data is publicly available, and metric evaluation on the validation set can be obtained through the CBICA imaging portal.
Apart from replacing the default \emph{DICE+BCE} loss with our custom-tailored loss candidates and, in some cases, necessary learning rate adjustments, we keep all training parameters constant.\footnote{Friendly note to beloved reviewers: Naturally, the code for our nnU-net training will be publicly available on GitHub for full reproducibility.}
To shed some light on the training variance, we also train another five times with dice\_bce baseline.

\begin{figure*}[hbtp]
    \centering
    \includegraphics[width=1.0\textwidth]{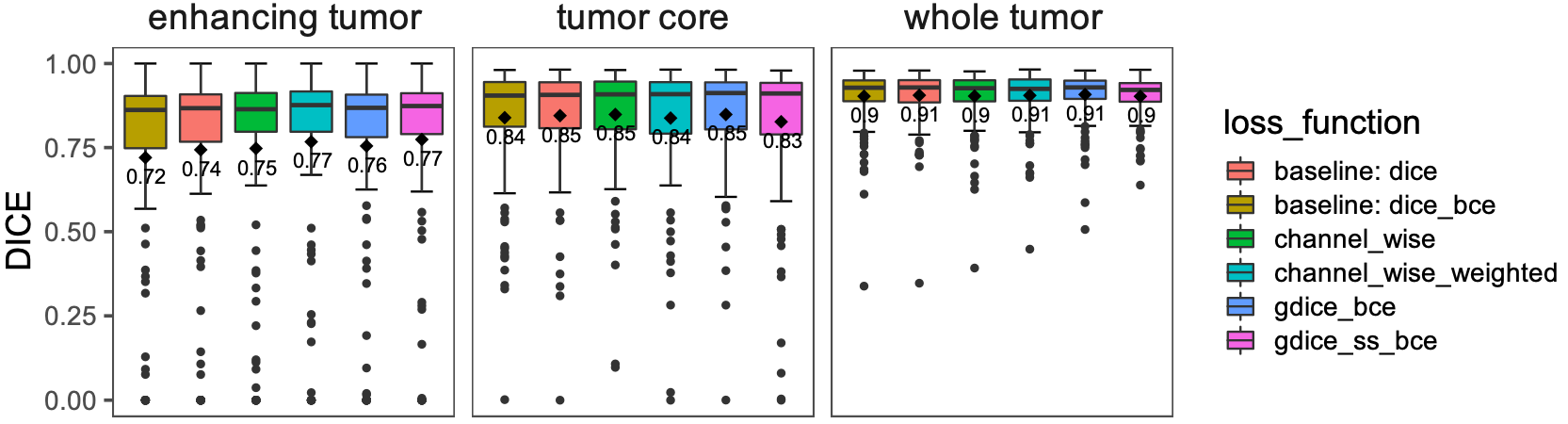}
    \caption{\small{Dice comparison of loss candidates vs. baselines across BraTS label channels.
    Diamonds indicate mean scores.
    P-values, without adjustment for repeated tests, for paired-samples T-tests comparing our candidates with the dice\_bce baseline from left to right: \emph{0.1547, 0.04642, 0.05958, 0.01881} for the 95\% significance level.
    While we consider the volumetric DICE score improvement as non-significant, interestingly, it is larger than the training variance, see Appendix.}}
    \label{fig:training_results}
\end{figure*}

\noindent \textbf{Results:}
We observe no significant quantitative improvement; the resulting volumetric dice scores are detailed in \Cref{fig:training_results}.

\subsection*{Experiment 6: Qualitative evaluation of compound losses}
\label{text:qualieva}
\noindent \textbf{Motivation:}
Even though we did not achieve a systematic improvement according to established segmentation quality metrics, we conduct another expert perception study to evaluate the generated compound losses with expert radiologists and fully close the loop.
As we learned from previous experiments that established quantitative metrics are not fully representative of expert assessment, we still consider this experiment worthy. 

\noindent \textbf{Procedure:}
Experts evaluate only the axial views of the gliomas' center of mass on a random sample of 50 patients from our test set.
We present our four candidate losses vs. the \emph{reference} and \emph{DICE+BCE} baseline as conditions.
This way, the experiment incorporates 300 trials.

\begin{figure}[H]
    \centering
    \includegraphics[width=0.95\textwidth]{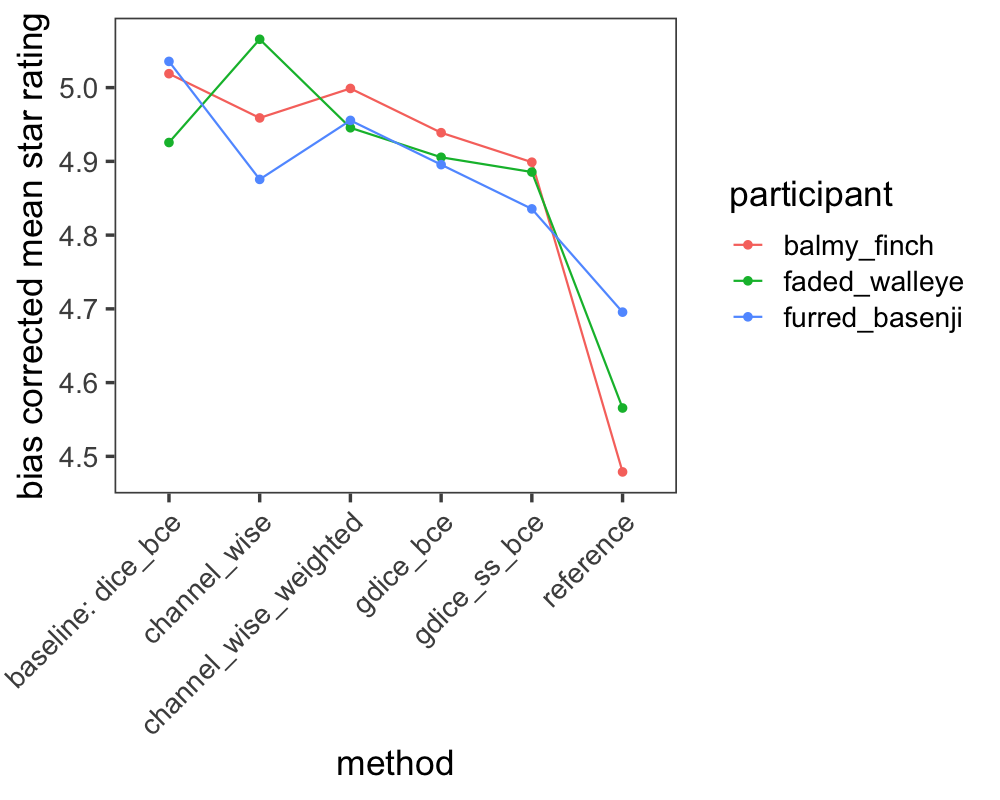}
    \caption{\small{
    Bias-corrected mean star ratings per expert and condition for the second MR segmentation rating experiment.
    Raw ratings and the bias correction methodology are detailed in the Appendix.
    Expert assessment of the four loss candidates vs. \emph{reference} and the \emph{DICE+BCE} baseline.
    We observe only subtle differences in expert rating between the \emph{DICE+BCE} baseline and our candidates.
    Notably, as in the first perception study, the human-annotated \emph{reference} is rated worse, compare \Cref{fig:brats_bc} \& \Cref{fig:brats_stars}.
    }} \label{fig:brats2_bc}
\end{figure}

\noindent \textbf{Results:}
Three male senior radiologists from two institutions participated in the experiment.
Participants reported three, five, and eleven years of work experience and were aged 31, 37, and 40 years. 
\Cref{fig:brats2_bc} illustrates how experts can hardly detect performance differences between the evaluated computer-generated segmentations.
Remarkably, again the human-annotated \emph{reference} is rated worse.

\section{Discussion}
We conducted multi-center expert perception studies with 15 expert radiologists for 3D MR- and eight expert radiologists for 2D ultrasound imaging segmentations.
Furthermore, to achieve a better fit with human expert assessment, we developed a method to construct compound loss functions for CNN training.
These compound loss functions represent an effort to better approximate abstract metrics that cannot be optimized directly, such as human expert assessment.
For this purpose, we select components based on the cluster - and principal component analysis and obtain weights from linear mixed models.
Even though we do not manage to train more expert-pleasing segmentation networks, our findings provide meaningful cues for future research.

Existing literature \citep{reinke2018exploit,maier2018rankings} reported on the discrepancy between expert segmentation quality assessment and established segmentation quality metrics.
Our approach, combining psychophysical experiments and classical statistics, identifies the quantitative correlates of qualitative human expert perception to shed more light on this complex phenomenon.
While both entities, \emph{segmentation quality metrics} and \emph{expert radiologists}, try to measure segmentation quality, our analysis reveals that their measurements correlate only moderately at best, especially for clinically relevant features such as the \emph{enhancing tumor} label.
Therefore, we highlight the need to develop new segmentation metrics and losses that better correlate with expert assessment.

One limitation of our study is that similarity metrics are always dependent on the employed reference annotations.
It is evident that \emph{DSC} fails to represent expert opinion when the segmentation of small structures matters, such as for multiple sclerosis lesion segmentation \citep{kofler2022blob} or metastasis segmentation \citep{buchner2022development}.
However, in the absence of actual ground truth, it is impossible to disentangle to which extent the observed low correlations are due to the incapability of metrics to capture expert assessment or a result of poor reference annotations \citep{kofler2022approaching}.
Nevertheless, it should be noted that the BraTS dataset arguably features above-average annotation quality, as, unlike other datasets in the biomedical field, it was curated by several experts over multiple years.

Another limitation of our study is that in the BraTS segmentation rating experiments, radiologists judge the complex 3D segmentations on three 2D views.
However, while axial views are rated slightly better, the rating effects are quite consistent across views (see \Cref{fig:per_view}).

When training a modern convolutional neural network with our expert-inspired compound loss functions, we cannot produce a significant improvement over the well-established \emph{Dice+BCE} baseline.
As we find \emph{BCE} to be one of few loss functions complementary to confusion-matrix-based losses and the \emph{Dice+BCE} baseline is mathematically similar to our identified loss candidates, this is perhaps not surprising (see \Cref{fig:loss_map}).
Hence our findings might explain why the empirically found \emph{Dice+BCE} provides a solid baseline across many segmentation tasks.

As algorithms and humans can usually generate satisfactory pre-operative glioma segmentations, poor segmentations are under-represented in our sample.
We hypothesize that if we replace our random sample from the BraTS test set with an equally distributed sample covering the full spectrum of segmentation quality, we might find higher correlations between established similarity metrics and human expert perception.
Up to the level of inter-rater agreement, it seems possible to train visually appealing segmentation networks using a plain soft Dice loss \citep{kofler2022approaching,kofler2022dqe}.
\footnote{This might also explain why surrogate models for human star rating reached a higher correlation with volumetric \emph{DSC} \citep{kofler2022dqe} on a dataset with a broader spectrum of segmentation quality.}

Interestingly, across all our experiments, experts consistently scored CNN segmentations better than the human-curated \emph{reference} labels.
This is not only true for sophisticated ensembles (see \Cref{fig:brats_bc}) but also for segmentations generated by single CNNs (compare \Cref{fig:us_bc}, \Cref{fig:brats2_bc}, \Cref{fig:per_view}).
This highlights the need to develop better annotation procedures.

We can only speculate why computers might produce better segmentations than human annotators.
In the annotation process, human annotators produce systematic and random errors.
With enough training data, a CNN can abstract and will only learn systematic errors represented in the training data.
However, training a CNN can eliminate random, non-systematic human annotation errors by averaging these out.

When interviewed in a single-blind setting, radiologists repeatedly attributed their judgments to the better tracing of contours \footnote{We find inconclusive multifaceted, moderate correlations between shape features of specific labels and expert ratings.}.
This seems plausible, as, unlike computers, humans do not possess unlimited diligence to pixel/voxel-perfectly trace contours.
Further, this is supported by the fact that the difference between algorithmic and human-annotated reference segmentations is least pronounced in the axial view \Cref{fig:per_view} that is the basis for most expert annotations (see \Cref{fig:per_view}).
However, how these complex decisions are formed remains an open research question.

\noindent\textbf{Conclusion:}
Even though the terms are often used synonymously in the literature, our experiments highlight that it is imperative to reflect upon whether \emph{reference} labels actually qualify as \emph{ground truth}.
In our experiments, a hypothetical volumetric DICE score of 1, so 100 percent overlap with the \emph{reference}, would translate to a worse segmentation quality, according to expert radiologists.
Therefore, our findings question the status of the Dice coefficient as the \emph{de facto gold standard} for measuring segmentation quality beyond expert agreement.
Furthermore, they question whether segmentation performance in general and challenge rankings, in particular, should be determined based on similarity metrics with potentially erroneous human-curated \emph{reference} annotations \citep{kofler2022approaching}.

Future research should investigate whether humans are still able to outperform CNNs when applying exhaustive quality assurance and control procedures.
While more participants would certainly add to the credibility of our experiments, we found the reported effects to be consistent across participants.
As senior radiologists are hard to come by, one should mention that this, to the best of our knowledge, represents the largest study featuring expert segmentation quality ratings.
As primate perception follows a fixed, partially understood set of rules, for instance, \citep{wagemans2012century,roelfsema2006cortical}, it might be that our findings also generalize to other segmentation problems, even outside the medical domain.
However, further research is required to explore this.

\clearpage


\acks{
\noindent Even though we cannot reveal their names to maintain anonymity, we want to thank the participating radiologists who enabled these studies.

\noindent Bjoern Menze, Benedikt Wiestler and Florian Kofler are supported through the SFB 824, subproject B12.

\noindent Supported by Deutsche Forschungsgemeinschaft (DFG) through TUM International Graduate School of Science and Engineering (IGSSE), GSC 81.

\noindent Suprosanna Shit and Ivan Ezhov are supported by the Translational Brain Imaging Training Network(TRABIT) under the European Union's `Horizon 2020' research \& innovation program (Grant agreement ID: 765148).

\noindent Ivan Ezhov, Suprosanna Shit are funded by DComEX (Grant agreement ID: 956201).

\noindent With the support of the Technical University of Munich – Institute for Advanced Study, funded by the German Excellence Initiative.

\noindent Supported by Anna Valentina Lioba Eleonora Claire Javid Mamasani.

\noindent Research reported in this publication was supported in part by the National Cancer Institute (NCI) of the National Institutes of Health, under award number U01CA242871. The content of this publication is solely the responsibility of the authors and does not represent the official views of the NIH.

\noindent Suprosanna Shit is supported by the Graduate School of Bioengineering,  Technical University of Munich.

\noindent Jan Kirschke has received Grants from the ERC, DFG, BMBF and is Co-Founder of Bonescreen GmbH.

\noindent Bjoern Menze acknowledges support by the Helmut Horten Foundation.

\noindent Research reported in this publication was partly supported by AIME GPU cloud services.
}


%
\ethics{The work follows appropriate ethical standards in conducting research and writing the manuscript, following all applicable laws and regulations regarding treatment of animals or human subjects.}

\coi{We declare we don't have conflicts of interest.}

\newpage

\bibliography{flow/meta/references.bib}


\clearpage
\appendix
\section{Terminology definitions}
Here we define some basic concepts of Neuroradiology and experimental Psychology used in the manuscript.

\noindent\textbf{Exam:} A radiology exam consists of a set of image recordings recorded in one session.
For instance, a glioma exam in the BraTS challenge consists of a T1, T1c, T2, and FLAIR image.

\noindent\textbf{Stimulus (material):} Unlike the control elements, which are constant over all trials, the stimulus material is varied across experimental conditions.
The participants' responses are recorded and analyzed with regard to different stimuli.

\noindent\textbf{Fixation cross:} In the inter-trial interval (ITI), we display fixation crosses to avoid confounding the recording of reaction times with \emph{unnecessary} eye movements such as \mbox{(re-)focusing} on the screen.

\section{Supplemental figures}
\begin{figure}[htbp]
    \centering
    \includegraphics[width=0.95\textwidth]{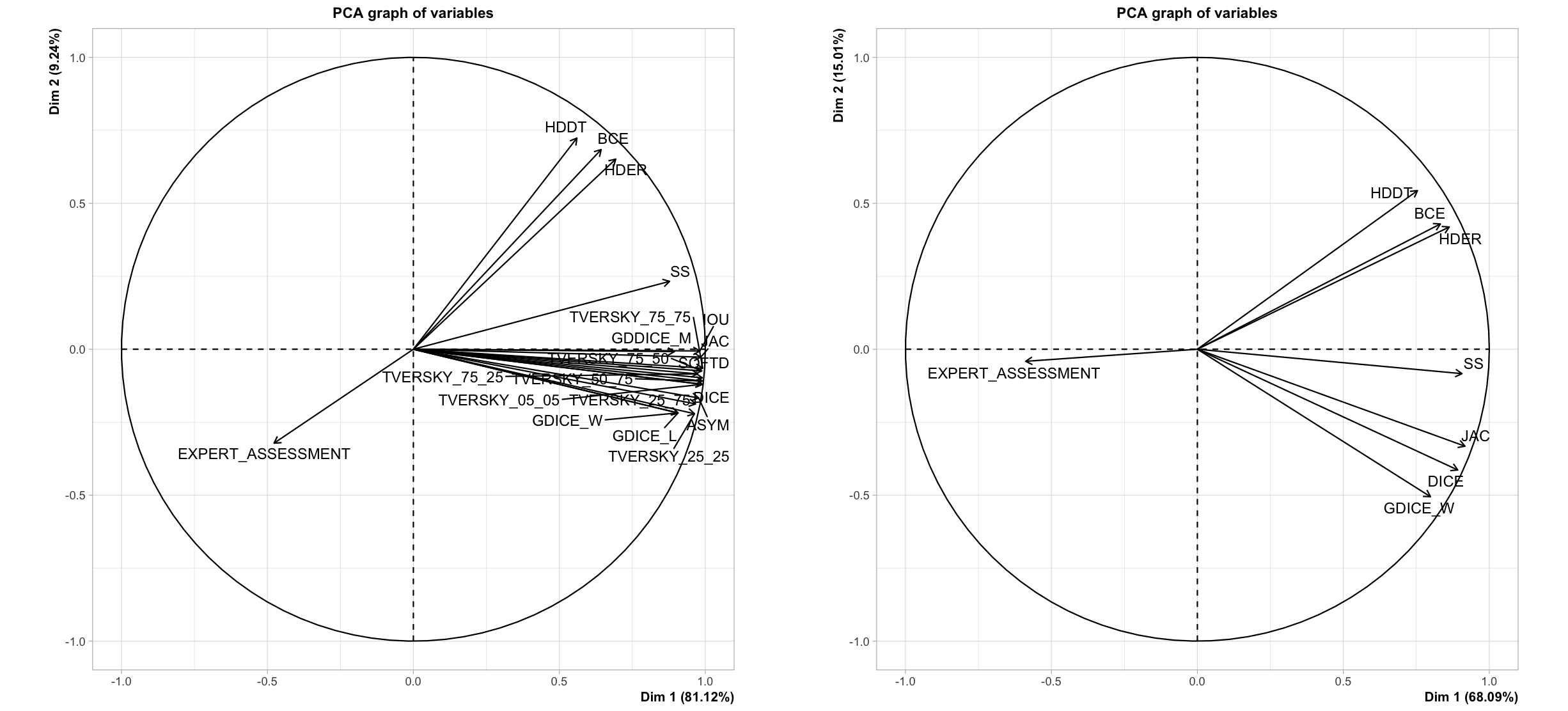}
    \caption{\small{PCA with two factors.
    The scree plot and multiple nongraphical tests show that two factors are sufficient to describe the loss landscape.
    PCA analysis with the full set of loss functions on the \emph{left}, for the sake of clarity, we repeated the PCA analysis with a subset of commonly used loss functions on the \emph{right}.
    Similar to the hierarchical clustering, one can see how expert assessment can be described as a combination of two factors.
    While the volumetric losses load mainly on dimension one, only \emph{BCE} and the two Hausdorff losses cover dimension two.
    This might explain why the empirically found baseline \emph{DICE+BCE} performs well across many segmentation tasks. }}
    \label{fig:pca}
\end{figure}

\begin{figure}[htbp]
\label{fig:per_view}
\centering
\includegraphics[width=0.95\textwidth]{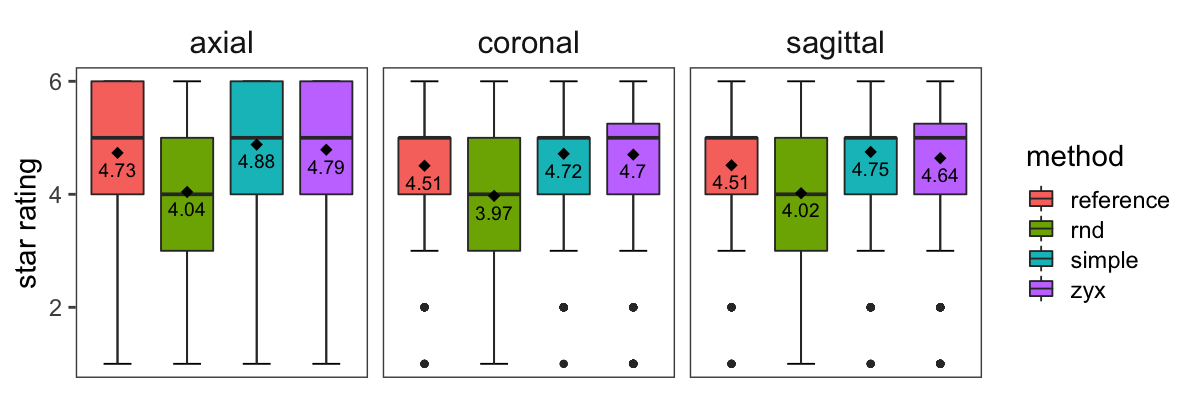}
\caption{\small{ Star rating across views. Diamonds indicate mean scores. The distribution is constant  per \emph{axial}, \emph{coronal} and \emph{sagital} view. In general, segmentation quality ratings seem slightly higher on the axial view. We speculate this might occur because doctors usually annotate on the axial view, so the resulting annotation quality should be slightly higher. In addition, the resolution is often higher in the axial plane leading to less ambiguity when judging the quality of segmentation images.}} 
\end{figure}

\begin{figure}[htbp]
\centering
\includegraphics[width=1.0\textwidth]{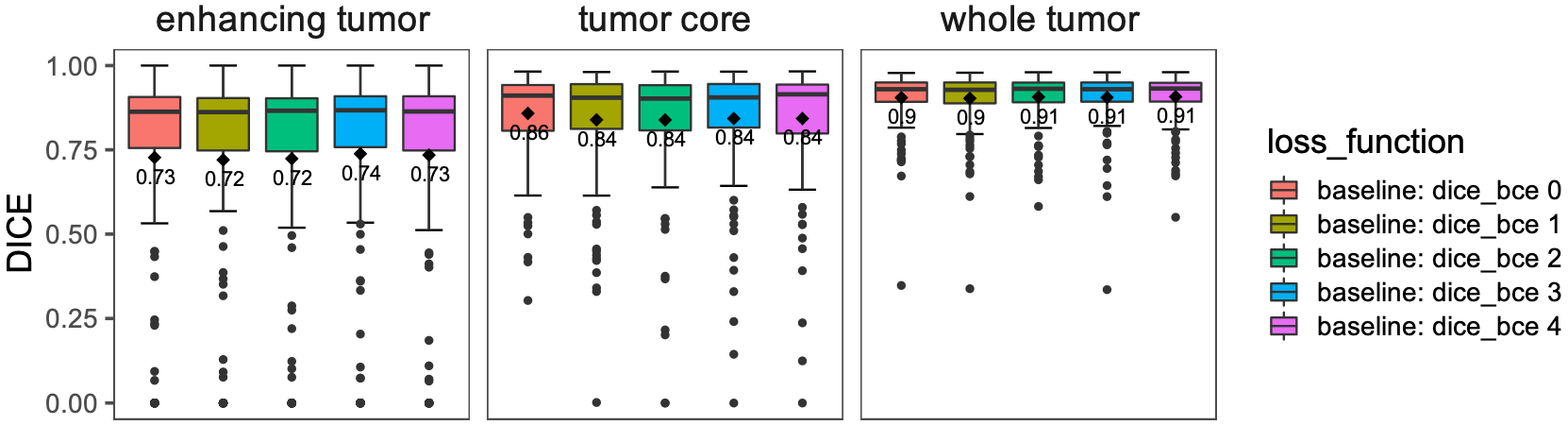}
\caption{\small{Illustration of training variance. Diamonds indicate mean scores. To get an estimate of the randomness involved in our training process, we trained our \emph{nnU-net} implementation five times with \emph{DICE+BCE} loss. We observe a dice performance between \emph{0.72} and \emph{0.74}, which is lower than the performance of our loss candidates.}} \label{fig:training_variance}
\end{figure}

\begin{figure}[htbp]
\centering
\includegraphics[width=1.0\textwidth]{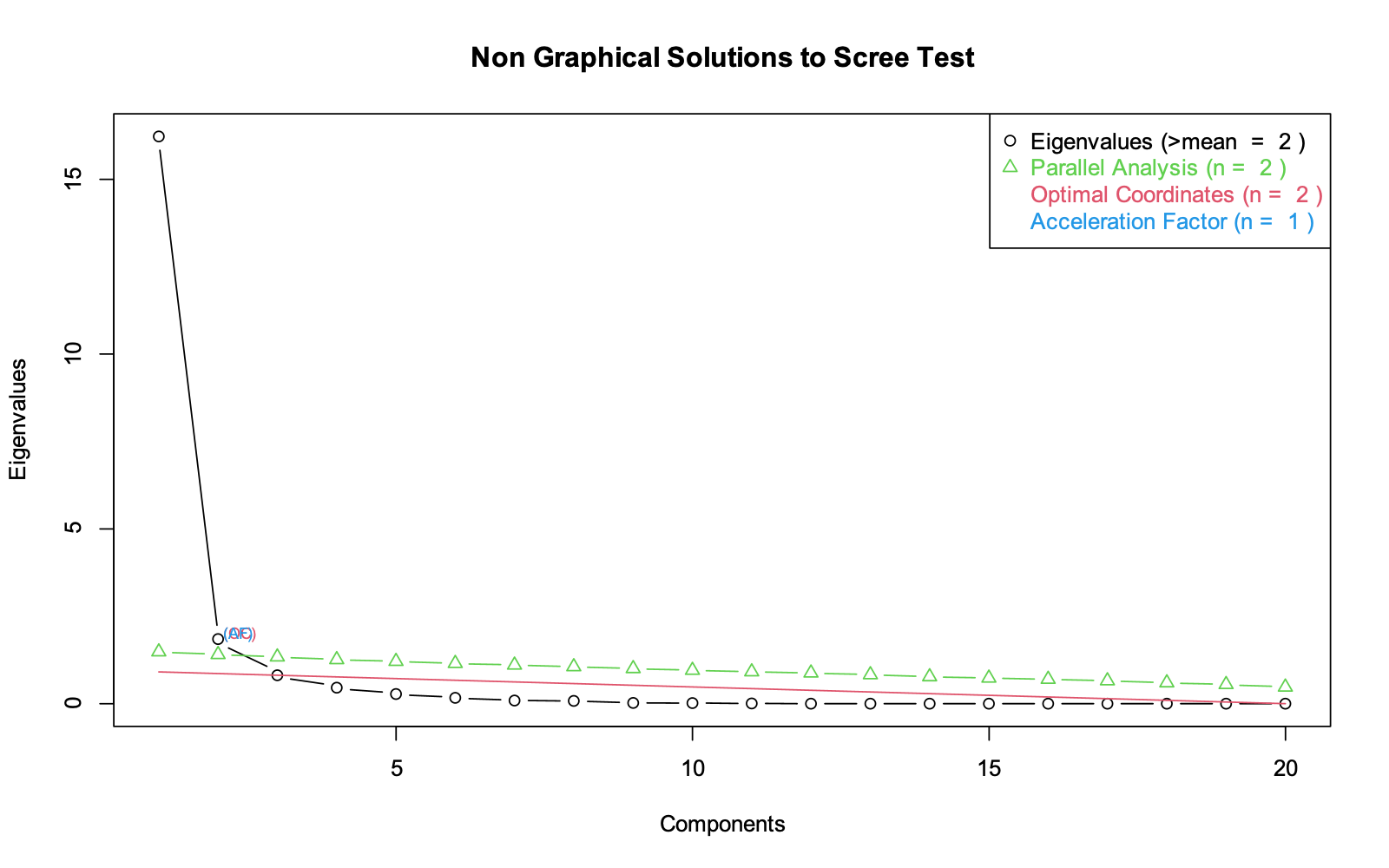}
\caption{\small{Scree plot. Multiple tests (see top right corner) indicate that two components are sufficient for the PCA.}} \label{fig:pca_scree}
\end{figure}

\begin{figure}[htbp]
    \centering
    \includegraphics[width=1.0\linewidth]{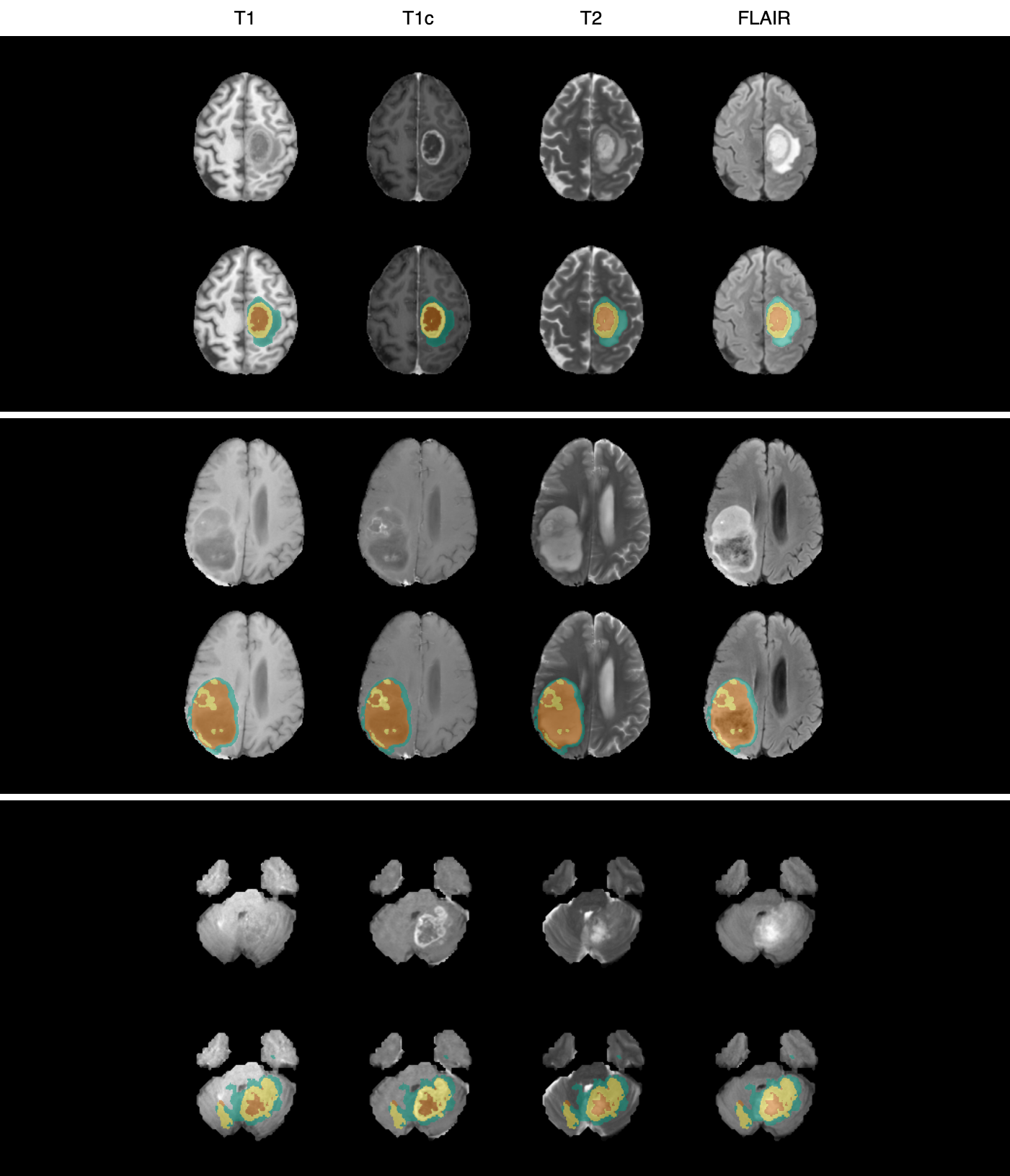}
    \caption{\small{Three segmentation examples with varying expert assessment.
    T1, T1c, T2, and FLAIR images are displayed from left to right.
    The segmentations are overlayed in \emph{red} for \emph{necrosis}, \emph{yellow} for \emph{enhancing tumor} and \emph{green} for \emph{edema}.
    The \emph{15} expert radiologists rated the top example \emph{superior} with a mean of \emph{5.73} stars, the middle example \emph{mediocre} with a mean of \emph{3.33} stars and the bottom example \emph{inferior} with a mean of \emph{1.93} stars.}} \label{fig:qualitative_examples}
\end{figure}


\clearpage
\section{Bias correction}

\begin{figure}[H]
    \centering    \includegraphics[width=0.55\linewidth]{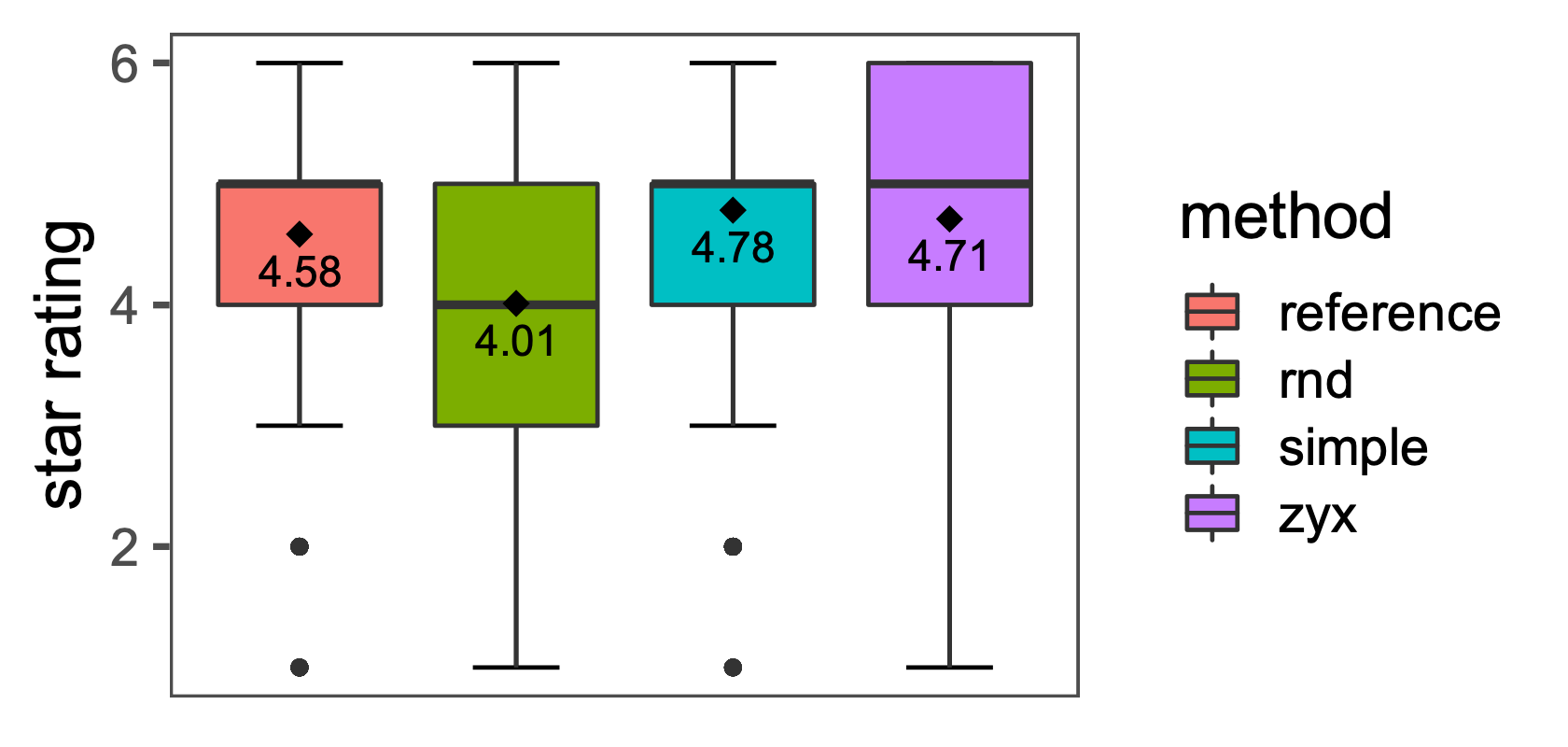}
    \caption{\small{Raw (non-averaged) expert assessment in star rating across exams for the different experimental conditions.
    Diamonds indicate mean scores.
    }} \label{fig:brats_raw_box}
\end{figure}

\begin{figure}[H]
    \centering
    \includegraphics[width=0.55\linewidth]{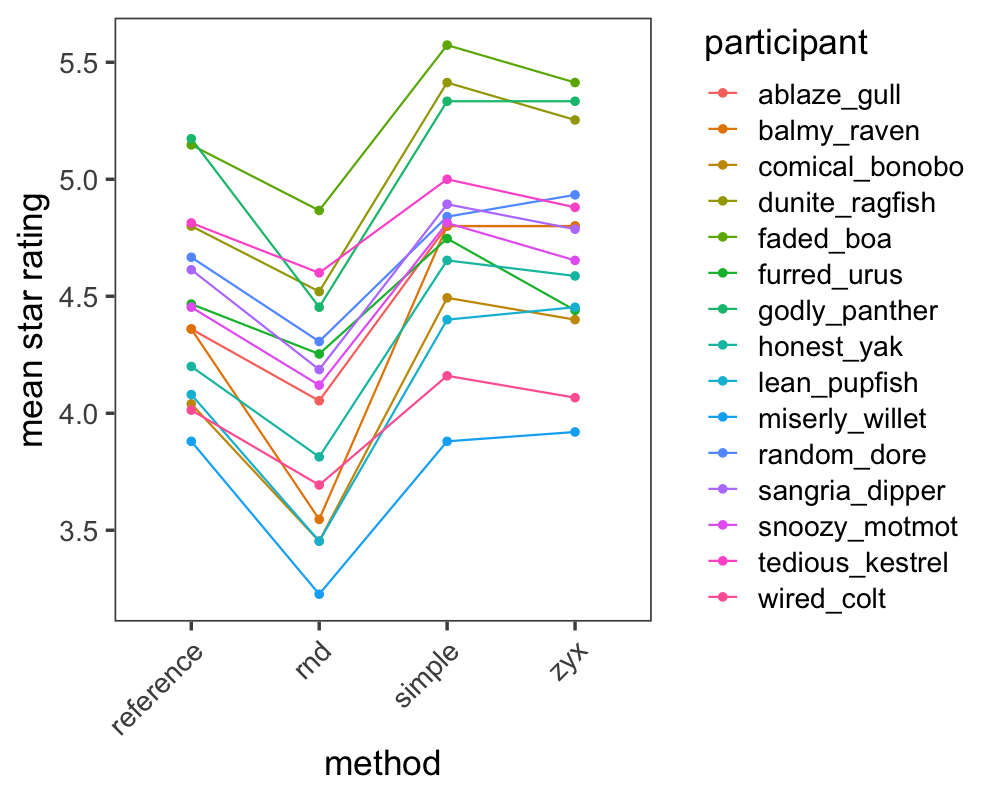}
    \caption{\small{
    Mean star ratings per participant and method for MR segmentation rating Experiment 1.
    Some participants reveal a strong / positive biases compared to the mean ratings, compare \Cref{fig:brats_biases}.
    However, when correcting for these biases participants are remarkably consistent in their ratings, compare \Cref{fig:brats_bc}.
    }} \label{fig:brats_raw}
\end{figure}

\begin{figure}[H]
    \centering
    \includegraphics[width=0.55\linewidth]{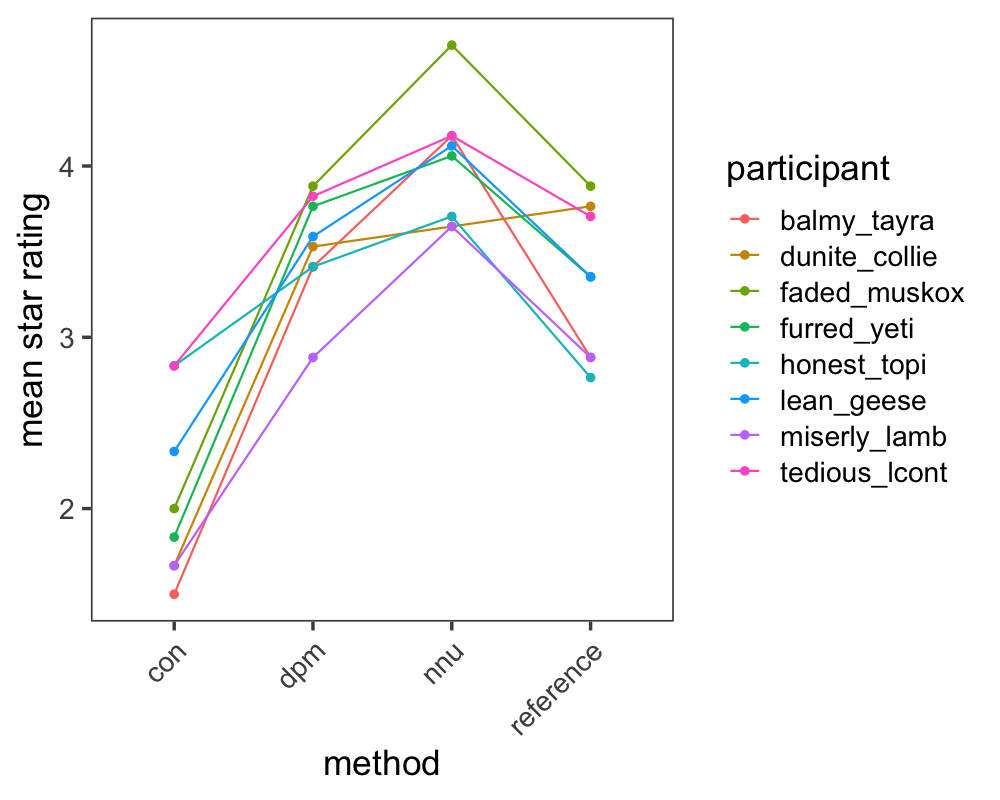}
    \caption{\small{
    Mean star ratings per participant and method for ultrasound segmentation rating experiment 2.
    Similar to the MR experiment, some participants reveal a strong / positive biases compared to the mean ratings.
    However, when correcting for these biases participants are remarkably consistent in their ratings, compare \Cref{fig:us_bc}.
    }} \label{fig:us_raw}
\end{figure}

\begin{figure}[H]
    \centering
    \includegraphics[width=0.65\textwidth]{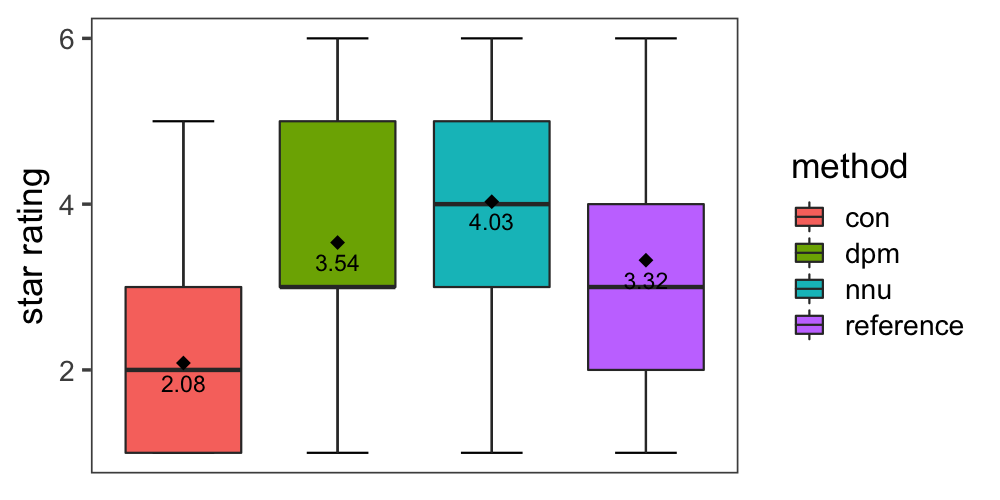}
    \caption{\small{
    Raw (non-averaged) expert assessment in star rating across exams in the ultrasound segmentation rating experiment 2.
    Diamonds indicate mean scores.
    Expert radiologists rated the \emph{nnu} (inspired by \citep{isensee2021nnu}) and \emph{dpm} (inspired by \citep{kamnitsas2017efficient,kamnitsas2015multi}) candidate algorithms much higher than the control condition (\emph{con}) consisting out of purposely wrongly manufactured segmentations.
    }} \label{fig:us_stars}
\end{figure}

\begin{figure}[H]
    \centering
    \includegraphics[width=0.65\linewidth]{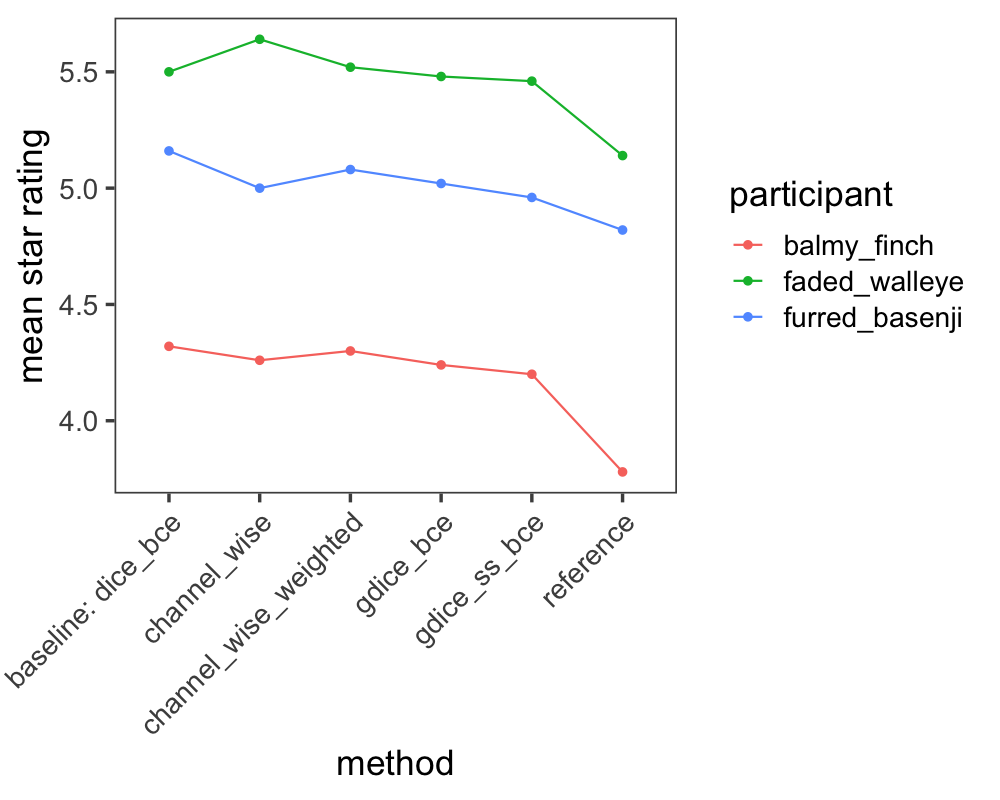}
    \caption{\small{
    Mean star ratings per participant and condition for MR segmentation rating Experiment 6.
    Some participants reveal strong biases in relation to the mean ratings.
    However, when correcting for these biases, participants are remarkably consistent in their ratings, compare \Cref{fig:brats2_bc}.
    }} \label{fig:brats2_raw}
\end{figure}


\begin{figure*}[hbtp]
    \centering
    \includegraphics[width=1.0\linewidth]{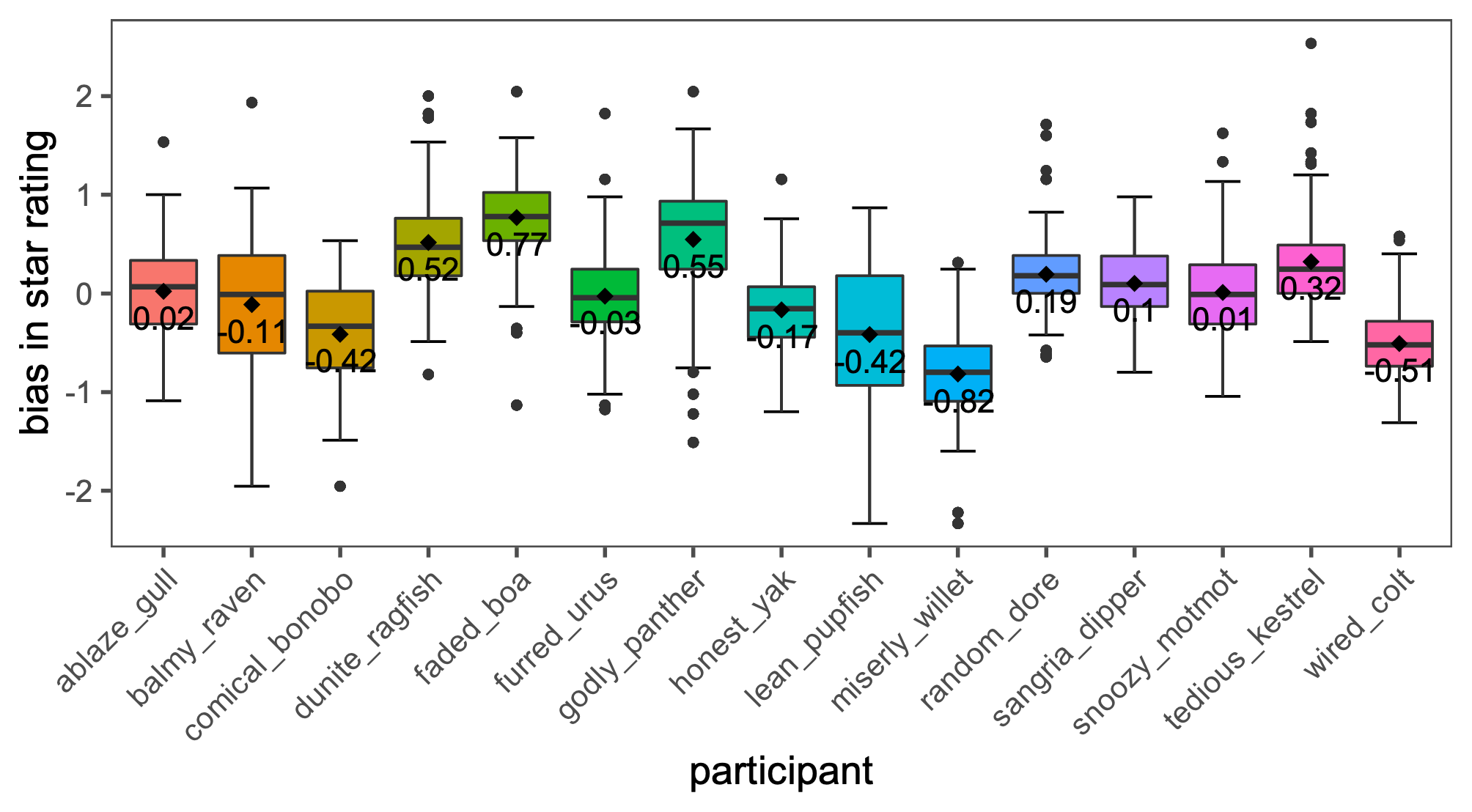}
    \caption{\small{Individual bias per participant in Experiment 1.
    Diamonds indicate the mean.
    The boxplots illustrate the difference between the participants individual ratings and the mean rating per exam accross all participants in Experiment 1.
    While some participants are very congruent with the average, e.g. participant \emph{ablaze\_gull}, we observe strong inter-individual differences for others like \emph{miserly\_willet}.
    One can obtain bias corrected ratings by subtracting the respective bias from the participants individual ratings resulting, compare \Cref{fig:brats_bc}.
    }} \label{fig:brats_biases}
\end{figure*}

\clearpage
\section{Linear mixed models}
\label{text:appendix_lmms}
Following, we illustrate the model outputs and model diagnostics for the linear mixed models derived in \Cref{text:losses_derived}.
The model diagnostics are generated with \emph{performance} \citep{ludecke2021performance}.
The model coefficients serve as weights for our loss functions. For the \emph{channel\_wise\_weighted} loss functions the more clinically relevant \emph{tumor core} and \emph{enhancing tumor} channel are weighted with a factor of \emph{five} over the \emph{whole tumor} channel for the channel wise aggregation.

Note: in the model outputs the \emph{segmentation method} is referred to as  \emph{condition}.
Further, the \emph{exam} is encoded as \emph{patient} (in our dataset there is only one exam per patient so the variables encode the same concept).

\newpage
\subsection{\emph{gdice\_bce} loss}
\lstset{basicstyle=\small\ttfamily,breaklines=true}
\lstset{framextopmargin=50pt,frame=bottomline}
\begin{lstlisting}[float=hbpt,frame=tb,caption=R model output for gdice\_bce loss,label=nf2]
Linear mixed model fit by REML.
t-tests use Satterthwaite's method ['lmerModLmerTest']
Formula: stars ~ BCE + GDICE_W + (1 | patient) + (1 | condition)
   Data: mm_losses_df

REML criterion at convergence: 106.9

Scaled residuals: 
    Min      1Q  Median      3Q     Max 
-1.7200 -0.3908  0.1478  0.4664  2.1067 

Random effects:
 Groups    Name        Variance Std.Dev.
 condition (Intercept) 0.11107  0.3333  
 patient   (Intercept) 0.21882  0.4678  
 Residual              0.09391  0.3065  
Number of obs: 72, groups:  condition, 26; patient, 24

Fixed effects:
            Estimate Std. Error      df t value Pr(>|t|)    
(Intercept)   4.0671     0.3146 65.7925  12.927  < 2e-16 ***
BCE          -0.4624     0.1024 50.4863  -4.514 3.83e-05 ***
GDICE_W      -0.7462     0.2896 58.8458  -2.577   0.0125 *  
---
Signif. codes:  0 '***' 0.001 '**' 0.01 '*' 0.05 '.' 0.1 ' ' 1

Correlation of Fixed Effects:
        (Intr) BCE   
BCE     -0.642       
GDICE_W  0.814 -0.306
           R2m       R2c
[1,] 0.3911532 0.8650808
\end{lstlisting}

\begin{figure}[htbp]
\centering
\includegraphics[width=0.98\textwidth]{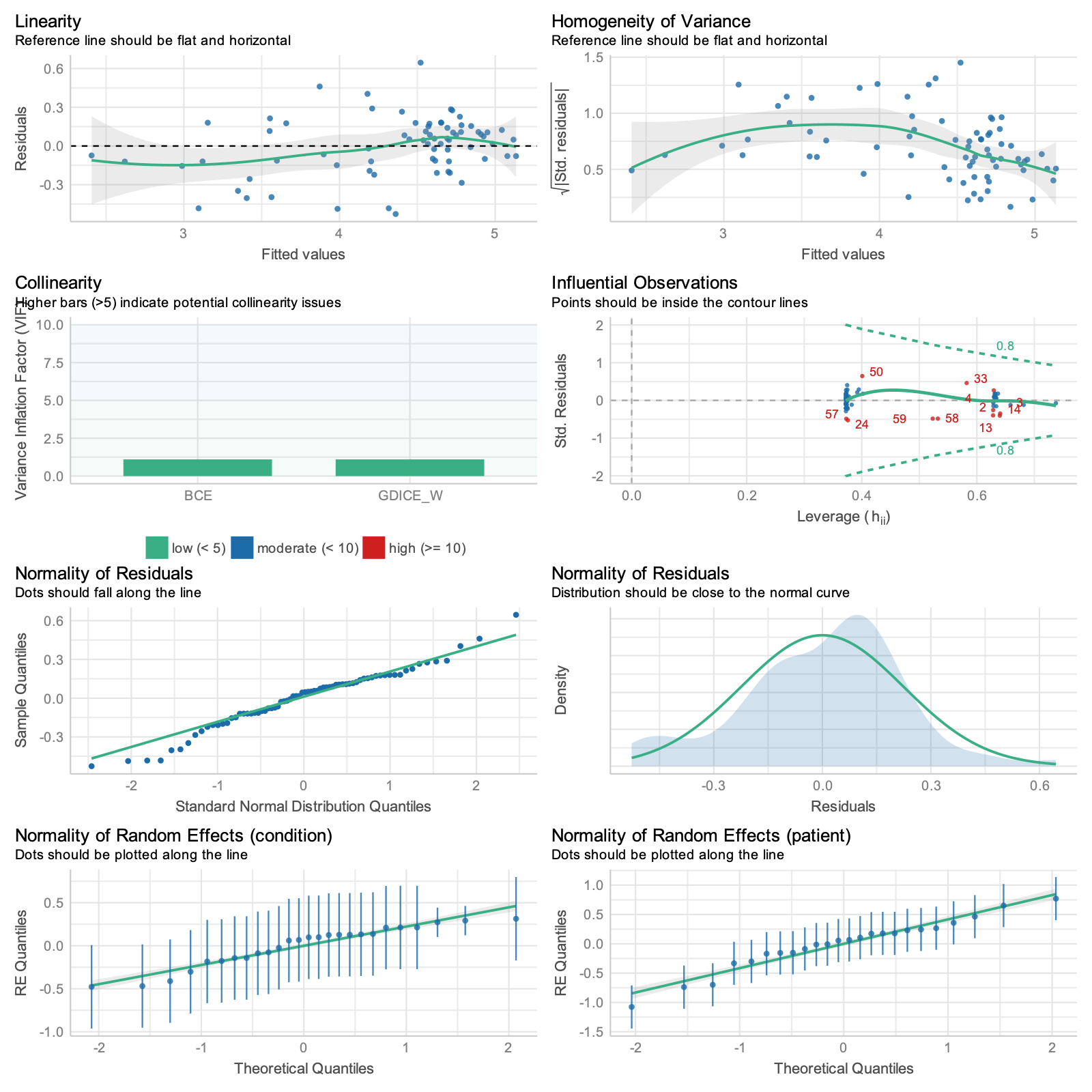}
\caption{
     \small{
     Model diagnostics for the \emph{gdice\_bce} loss.
     The loss function ombined \emph{BCE} and \emph{GDICE\_W} with weights \emph{0.4624} and \emph{0.7462} obtained from the respective coefficients in the linear mixed model.
     The model achieved a \emph{Pseudo R2} of \emph{0.391}.
     )}
}
\label{fig:model_gdicebce}
\end{figure}

\clearpage
\subsection{\emph{gdice\_bce\_ss} loss}
\lstset{basicstyle=\small\ttfamily,breaklines=true}
\lstset{framextopmargin=50pt,frame=bottomline}
\begin{lstlisting}[float=hbpt,frame=tb,caption=R model output for gdice\_bce\_ss loss,label=nf3]
Linear mixed model fit by REML.
t-tests use Satterthwaite's method [
lmerModLmerTest]
Formula: stars ~ BCE + GDICE_W + SS + (1 | patient) + (1 | condition)
   Data: mm_losses_df

REML criterion at convergence: 98.3

Scaled residuals: 
    Min      1Q  Median      3Q     Max 
-1.9685 -0.4159  0.1278  0.4612  2.1035 

Random effects:
 Groups    Name        Variance Std.Dev.
 condition (Intercept) 0.11433  0.3381  
 patient   (Intercept) 0.25798  0.5079  
 Residual              0.08433  0.2904  
Number of obs: 72, groups:  condition, 26; patient, 24

Fixed effects:
            Estimate Std. Error       df t value Pr(>|t|)    
(Intercept)   4.4306     0.3829  63.7986  11.570   <2e-16 ***
BCE          -0.3267     0.1693  42.6522  -1.930   0.0603 .  
GDICE_W      -0.4570     0.3439  50.1523  -1.329   0.1899    
SS          -18.2016    13.9181  54.4509  -1.308   0.1964    
---
Signif. codes:  0 '***' 0.001 '**' 0.01 '*' 0.05 '.' 0.1 ' ' 1

Correlation of Fixed Effects:
        (Intr) BCE    GDICE_
BCE      0.125              
GDICE_W  0.863  0.289       
SS      -0.569 -0.788 -0.556
         R2m       R2c
[1,] 0.40454 0.8900317
\end{lstlisting}

\begin{figure}[htbp]
\centering
\includegraphics[width=0.98\textwidth]{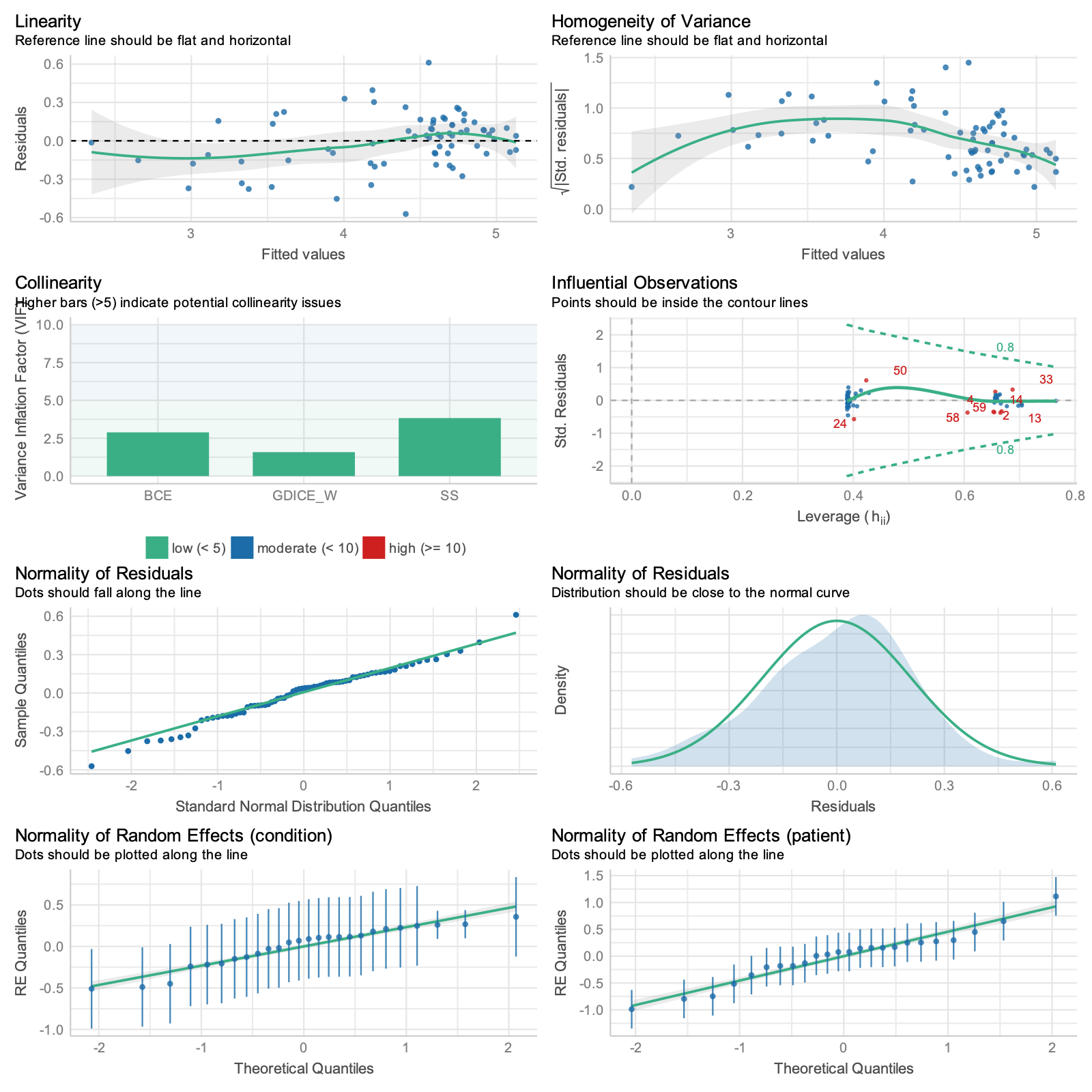}
\caption{
     \small{
     Model diagnostics for the \emph{gdice\_bce\_ss} loss.
     The loss function combined \emph{BCE}, \emph{GDICE\_W} and \emph{SS} with weights \emph{0.3267}, \emph{0.4570} and \emph{18.2016} obtained from the respective coefficients in the linear mixed model.
     The model achieved a \emph{Pseudo R2} of \emph{0.405}.
     }
}
\label{fig:model_gdicebcess}
\end{figure}

\clearpage
\subsection{channel wise losses: whole tumor channel}
\lstset{basicstyle=\small\ttfamily,breaklines=true}
\lstset{framextopmargin=50pt,frame=bottomline}
\begin{lstlisting}[float=hbpt,frame=tb,caption=R model output for whole tumor channel,label=cwwt]
Linear mixed model fit by REML. t-tests use Satterthwaite's method ['lmerModLmerTest']
Formula: stars ~ GDICE_W + SS + BCE + (1 | patient) + (1 | condition)
   Data: wt

REML criterion at convergence: 315.2

Scaled residuals: 
     Min       1Q   Median       3Q      Max 
-2.21560 -0.57743  0.00251  0.61567  2.83251 

Random effects:
 Groups    Name        Variance Std.Dev.
 condition (Intercept) 0.2637   0.5135  
 patient   (Intercept) 0.1590   0.3988  
 Residual              0.1630   0.4037  
Number of obs: 216, groups:  condition, 26; patient, 24

Fixed effects:
            Estimate Std. Error       df t value Pr(>|t|)    
(Intercept)   3.1165     0.9290 207.7590   3.355 0.000944 ***
GDICE_W      -1.5876     0.9146 204.4111  -1.736 0.084114 .  
SS           -4.0027    10.2444 196.2805  -0.391 0.696425    
BCE          -0.3039     0.0785 158.8889  -3.872 0.000157 ***
---
Signif. codes:  0 '***' 0.001 '**' 0.01 '*' 0.05 '.' 0.1 ' ' 1

Correlation of Fixed Effects:
        (Intr) GDICE_ SS    
GDICE_W  0.985              
SS      -0.893 -0.899       
BCE      0.492  0.545 -0.729
\end{lstlisting}

\begin{figure}[htbp]
     \centering
     \includegraphics[width=0.98\textwidth]{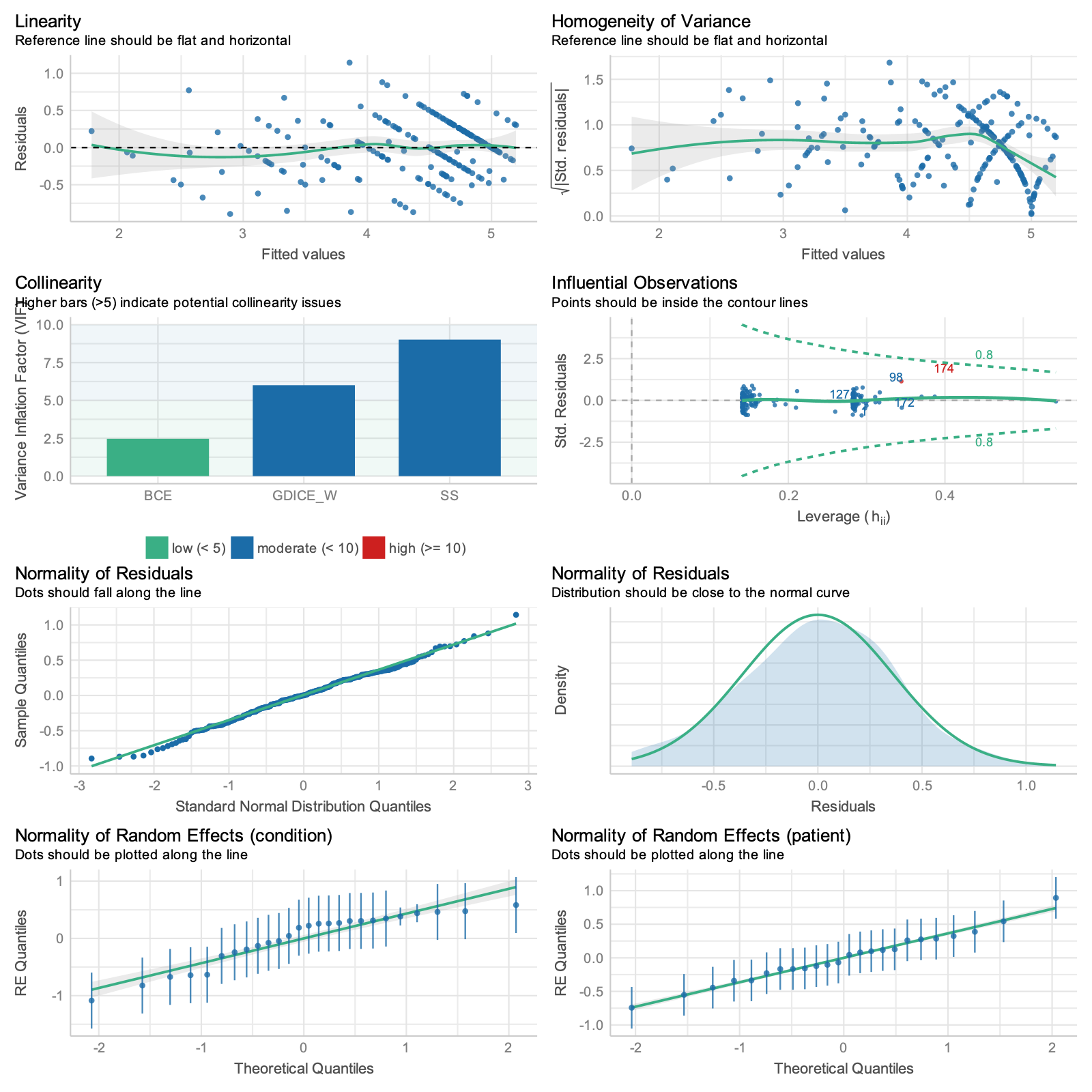}
     \caption{
          \small{
          Model diagnostics for the \emph{whole tumor} channels of the \emph{channel\_wise} and \emph{channel\_wise\_weighted} loss functions.
          The loss functions combined \emph{GDICE\_W},  \emph{SS} and \emph{BCE} with weights \emph{1.5876}, \emph{4.0027} and \emph{0.3039} obtained from the respective coefficients in the linear mixed model.
          Notably, we observed higher variation inflation than for \emph{gdice\_bce\_ss} loss combining the same components over all channels.
          The model achieved a \emph{Pseudo R2} of \emph{0.270}.
          }
     }
\label{fig:wt_model}
\end{figure}

\clearpage
\subsection{channel wise losses: tumor core channel}
\lstset{basicstyle=\small\ttfamily,breaklines=true}
\lstset{framextopmargin=50pt,frame=bottomline}
\begin{lstlisting}[float=hbpt,frame=tb,caption=R model output for tumor core channel,label=wctc]
Linear mixed model fit by REML.
t-tests use Satterthwaite's method ['lmerModLmerTest']
Formula: stars ~ GDICE_W + BCE + (1 | patient) + (1 | condition)
   Data: tc

REML criterion at convergence: 366.6

Scaled residuals: 
    Min      1Q  Median      3Q     Max 
-2.8922 -0.6517  0.0618  0.6587  3.1648 

Random effects:
 Groups    Name        Variance Std.Dev.
 condition (Intercept) 0.1892   0.4349  
 patient   (Intercept) 0.1852   0.4303  
 Residual              0.2112   0.4596  
Number of obs: 216, groups:  condition, 26; patient, 24

Fixed effects:
            Estimate Std. Error       df t value Pr(>|t|)    
(Intercept)  3.76303    0.25975 19.38353  14.487 7.44e-12 ***
GDICE_W     -0.77646    0.26230 12.01047  -2.960   0.0119 *  
BCE         -0.21026    0.04496 40.97802  -4.676 3.16e-05 ***
---
Signif. codes:  0 '***' 0.001 '**' 0.01 '*' 0.05 '.' 0.1 ' ' 1

Correlation of Fixed Effects:
        (Intr) GDICE_
GDICE_W  0.817       
BCE     -0.492 -0.282
\end{lstlisting}

\begin{figure}[htbp]
\centering
\includegraphics[width=0.98\textwidth]{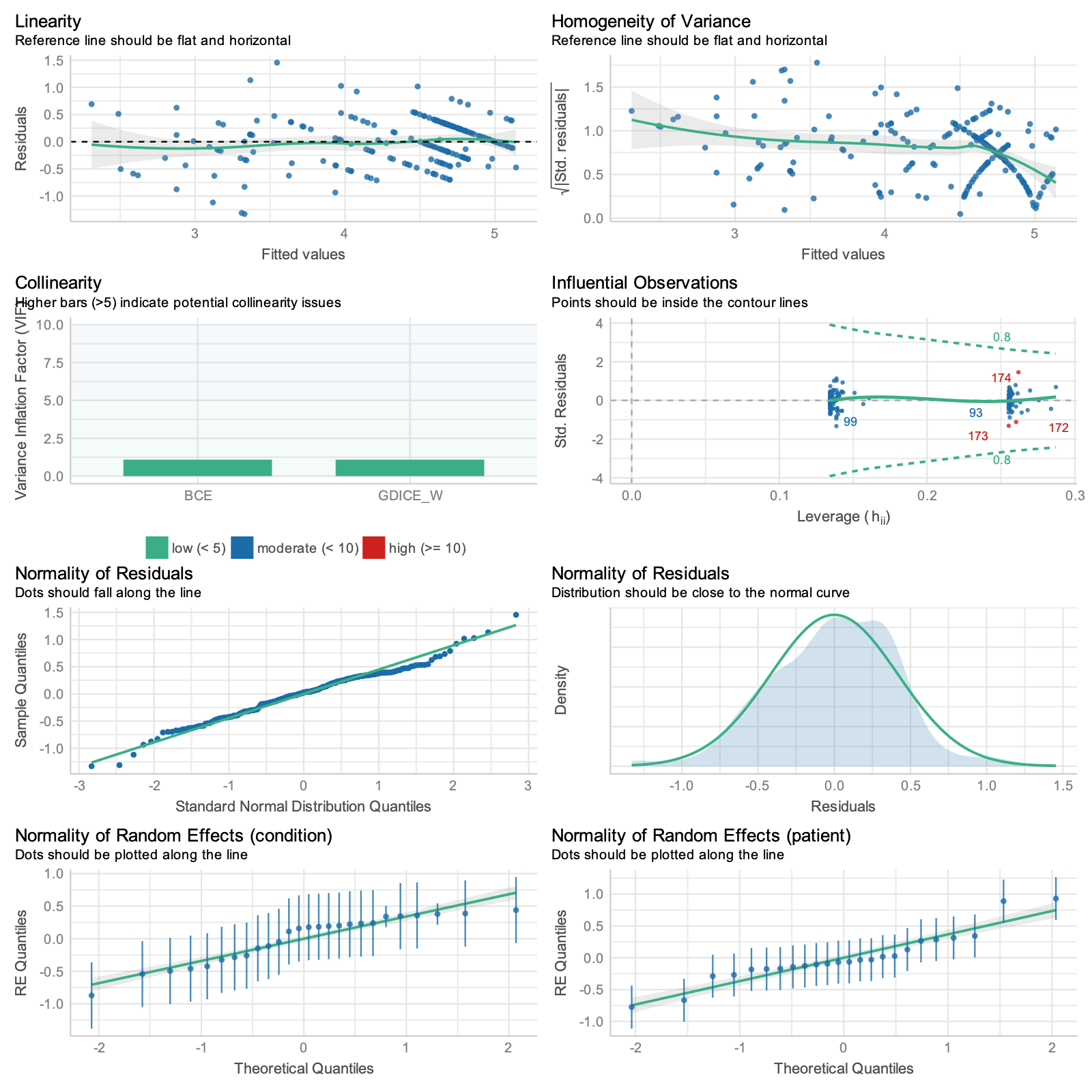}
\caption{
     \small{
     Model diagnostics for the \emph{tumor core} channels of the \emph{channel\_wise} and \emph{channel\_wise\_weighted} loss functions.
     The loss functions combined \emph{GDICE\_W} and \emph{BCE} with weights \emph{0.77646} and \emph{21026} obtained from the respective coefficients in the linear mixed model.
     The model achieved a \emph{Pseudo R2} of \emph{0.347}.
     }
}
\label{fig:tc_model}
\end{figure}

\clearpage
\subsection{channel wise losses: enhancing tumor channel}
\lstset{basicstyle=\small\ttfamily,breaklines=true}
\lstset{framextopmargin=50pt,frame=bottomline}
\begin{lstlisting}[float=hbpt,frame=tb,caption=R model output for enhancing tumor channel,label=cwet]
Linear mixed model fit by REML.
t-tests use Satterthwaite's method ['lmerModLmerTest']
Formula: stars ~ GDICE_W + (1 | patient) + (1 | condition)
   Data: et

REML criterion at convergence: 383.5

Scaled residuals: 
    Min      1Q  Median      3Q     Max 
-2.8596 -0.6098  0.0881  0.6515  3.6149 

Random effects:
 Groups    Name        Variance Std.Dev.
 condition (Intercept) 0.3716   0.6096  
 patient   (Intercept) 0.1203   0.3469  
 Residual              0.2308   0.4805  
Number of obs: 216, groups:  condition, 26; patient, 24

Fixed effects:
            Estimate Std. Error      df t value Pr(>|t|)    
(Intercept)   3.3635     0.2169 42.8141  15.509  < 2e-16 ***
GDICE_W      -0.9002     0.2318 26.7340  -3.884 0.000609 ***
---
Signif. codes:  0 '***' 0.001 '**' 0.01 '*' 0.05 '.' 0.1 ' ' 1

Correlation of Fixed Effects:
        (Intr)
GDICE_W 0.729 
\end{lstlisting}

\begin{figure}[htbp]
\centering
\includegraphics[width=0.98\textwidth]{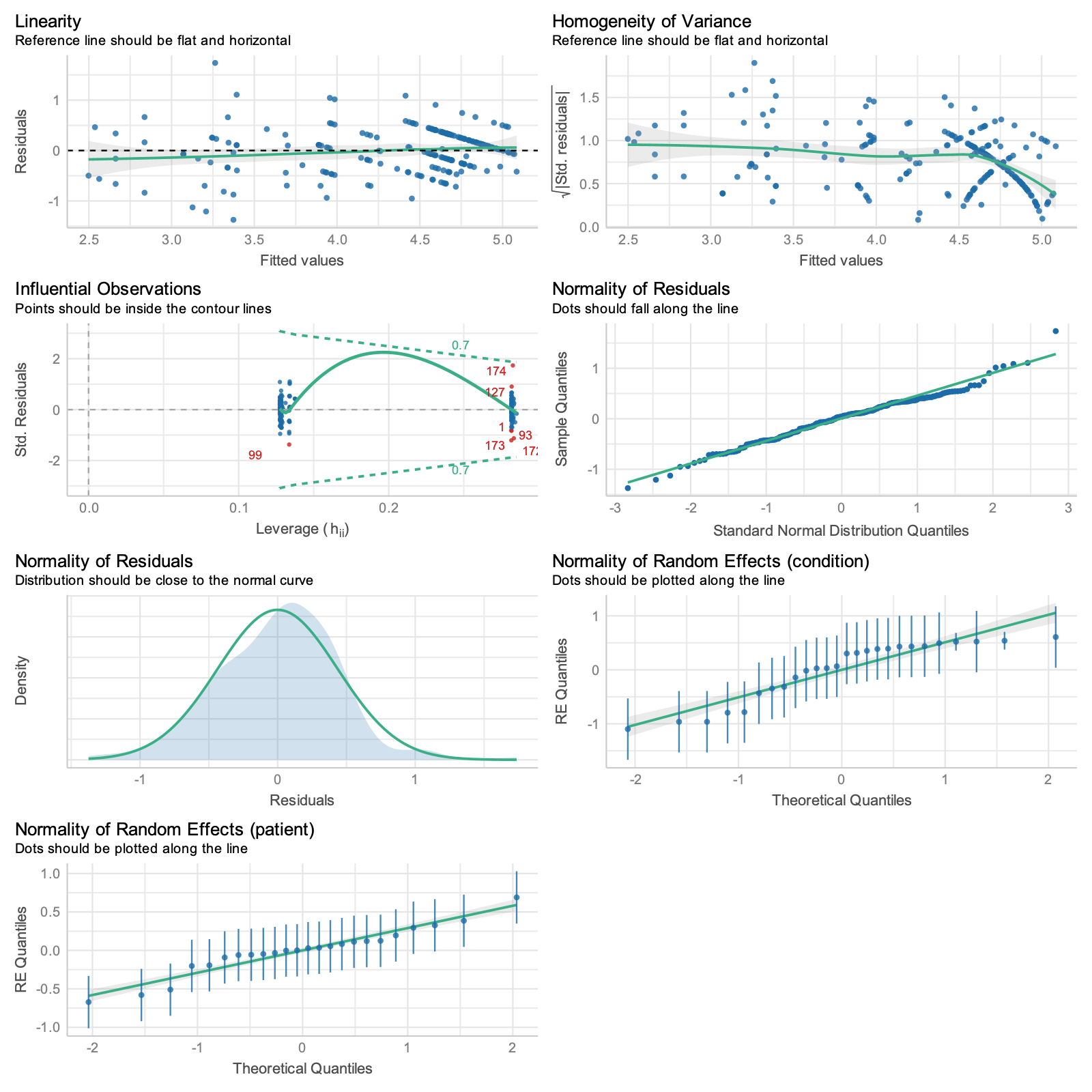}
\caption{
     \small{
     Model diagnostics for the \emph{enhancing tumor} channels of the \emph{channel\_wise} and \emph{channel\_wise\_weighted} loss functions.
     The loss functions combined \emph{GDICE\_W} and \emph{BCE} with weights \emph{0.77646} and \emph{21026} obtained from the respective coefficients in the linear mixed model.
     The model achieved a \emph{Pseudo R2} of \emph{0.112}, which is not surprising as our previous analysis showed that the clinically relevant \emph{enhancing tumor} label is very difficult to capture.
     }
}
\end{figure}

\end{document}